\documentclass[preprint2]{aastex62}

\usepackage[T1]{fontenc}
\usepackage{graphicx}
\usepackage{url}
\usepackage{amsmath}
\usepackage[T1]{fontenc}
\usepackage{aecompl}
\usepackage{color}
\usepackage{booktabs}
\usepackage[capitalize]{cleveref}

\graphicspath{{./fig/}}

\newcommand{\vsini}{\ensuremath{v \sin i}}
\newcommand{\Kepler}{\mbox{\textit{Kepler}}}

\newcommand{\Gaia}{\mbox{\textit{Gaia}}}

\newcommand{\Teff}{\ensuremath{T_{\textrm{eff}}}}
\newcommand{\logg}{\ensuremath{\log(g)}}
\newcommand{\kms}{\textrm{km~s}\ensuremath{^{-1}}}
\newcommand{\Ks}{\ensuremath{Ks}}
\newcommand{\MK}{\ensuremath{M_{\Ks}}}
\newcommand{\feh}{\textrm{[Fe/H]}}
\newcommand{\afe}{\textrm{[\ensuremath{\alpha}/Fe]}}
\newcommand{\mgfe}{\textrm{[Mg/Fe]}}

\newcommand{\STARBAD}{\texttt{STAR\_BAD}}
\newcommand{\STARWARN}{\texttt{STAR\_WARN}}

\newcommand{\APOGEECOOLDWARF}{\texttt{APOGEE\_KEPLER\_COOLDWARF}}
\newcommand{\APOKASC}{\texttt{APOGEE2\_APOKASC}}
\newcommand{\APOKASCDWARF}{\texttt{APOGEE2\_APOKASC\_DWARF}}
\newcommand{\APOKASCGIANT}{\texttt{APOGEE2\_APOKASC\_GIANT}}

\newcommand{\APOGEEKOI}{\texttt{APOGEE2\_KOI}}
\newcommand{\APOGEEKOICONTROL}{\texttt{APOGEE2\_KOI\_CONTROL}}
\newcommand{\APOGEESEISMO}{\texttt{APOGEE\_KEPLER\_SEISMO}}
\newcommand{\APOGEERVMONITOR}{\texttt{APOGEE\_RV\_MONITOR\_KEPLER}}

\newcommand{\APOGEEHOST}{\texttt{APOGEE\_KEPLER\_HOST}}

\shorttitle{\Kepler{} Subgiant Rotation}
\shortauthors{Simonian et al.}
\begin{document}

\title{Rapid Rotation of \Kepler{} field dwarfs and subgiants: Spectroscopic 
    \vsini{} from APOGEE}
\author[0000-0002-4230-6732]{Gregory V. A. Simonian}
\affiliation{Department of Physical Science, Concord University, 1000 Vermillion
    Street, Athens, WV 24712}
\affiliation{Department of Astronomy, The Ohio State University, 140 West 18th Avenue, Columbus, OH 43210}

\author[0000-0002-7549-7766]{Marc H. Pinsonneault}
\affiliation{Department of Astronomy, The Ohio State University, 140 West 18th Avenue, Columbus, OH 43210}

\author[0000-0002-0431-1645]{Donald M. Terndrup}
\affiliation{Department of Astronomy, The Ohio State University, 140 West 18th Avenue, Columbus, OH 43210}

\author[0000-0002-4284-8638]{Jennifer L. van Saders}
\affiliation{Institute for Astronomy University of Hawai'i 
Honolulu, HI 96822}
\affiliation{Observatories of the Carnegie Institution for Science 
Pasadena, CA 91101}
\affiliation{Department of Astrophysical Sciences, Princeton University 
Princeton, NJ 08544}

\correspondingauthor{Gregory V. A. Simonian}
\email{gsimonian@concord.edu}

\begin{abstract}
  We use 5,337 spectroscopic \vsini{} measurements of \Kepler{} dwarfs and
  subgiants from the APOGEE survey to study stellar rotation trends. We 
  find a detection threshold of 10 km/s, which allows us to explore
  the spindown of intermediate-mass stars leaving the main sequence, merger
  products, young stars, and tidally-synchronized binaries. We see a clear 
  distinction between blue stragglers and the field turnoff in \(\alpha\)-rich 
  stars, with a sharp rapid rotation cutoff for blue stragglers consistent with 
  the Kraft break. We also find rapid rotation and RV variability in a sample 
  of red straggler stars, considerably cooler than the giant branch, lending 
  credence to the hypothesis that these are active, tidally-synchronized 
  binaries. We see clear evidence for a transition between rapid and slow 
  rotation on the subgiant branch in the domain predicted by modern angular 
  momentum evolution models. We find substantial agreement between the 
  spectroscopic and photometric properties of KIC targets added by 
  \citet{Huber14} based on 2MASS photometry. For the unevolved lower main 
  sequence, we see the same concentration toward rapid rotation in photometric 
  binaries as that observed in rotation period data, but at an enhanced rate. 
  We attribute this difference to unresolved near-equal luminosity 
  spectroscopic binaries with velocity displacements on the order of the 
  APOGEE resolution. Among cool unevolved stars we find an excess rapid 
  rotator fraction of 4\% caused by pipeline issues with photometric binaries.
\end{abstract}

\keywords{stars: rotation, stars: late-type}

\section{Introduction}

Rotation is a fundamental property of stars. We observe a wide range of
rotation rates in pre-main sequence stars of all masses \citep{Attridge92,
Herbst00, Henderson12}. \citet{Kraft67} noticed a dichotomy in rotation rates
for stars on the main sequence, with stars more massive than \(1.2 M_\sun{}\) maintaining a wide range of
rotation rates, while less massive stars stars rotate substantially slower. The
transition between the two is thought to be because of the onset of convective
envelopes in lower-mass stars, which lose angular momentum through magnetized
winds \citep{Parker58,Weber67,Kawaler88}. Studying the transition between the
two regimes can be a test of wind theory \citep{vanSaders13}.

Rotation is valuable for understanding stellar physics not just because it can
tell us about stellar dynamos and winds, but also because rotation may yield 
inferences of stellar ages, a
technique known as gyrochronology \citep{Barnes07,Mamajek08}.  Gyrochronology
relations have been calibrated using rotation data from open clusters to infer 
ages from rotation rates and masses. However, open
cluster rotation measurements are primarily in young systems, where amplitudes
are high and periods are short, with few old calibrators available. Attempts to
calibrate the relations using \Kepler{} stars with 
asteroseismic ages revealed challenges in extending existing gyrochronology
relationships. \citet{Angus15} found that traditional empirical gyrochronology
relationships could not be simultaneously fit to young clusters, asteroseismic
targets and field stars. A more sophisticated treatment with theoretical models
of angular momentum evolution found that \Kepler{} asteroseismic stars did not
slow down as quickly as predicted by angular momentum models
\citep{vanSaders16}. Coincident studies of magnetic activity indicators among
\Kepler{} asteroseismic stars show that while the rotation of older stars
levels off, the magnetic field continues to decrease, suggesting a fundamental
change in the magnetic braking mechanism \citep{Metcalfe16,vanSaders16,
Metcalfe17}.

Although main-sequence gyrochronology is powerful, there are many other
astrophysical problems tied to the study of stellar rotation. Asteroseismology
has yielded crucial data on internal stellar rotation in evolved subgiants
\citep[e.g.][]{Deheuvels14} and red giants \citep{Mosser12,Gehan18} stars.
Measurements of surface rotation rates are important for interpreting the
subgiant core rotation data. The advent of
\Gaia{} parallaxes allows discrimination between dwarfs and subgiants in the
\Kepler{} field \citep{Berger18b}. 

Close binary stars can be tidally synchronized, and in an earlier paper we
found that the short rotation period systems in the \Kepler{} lower main
sequence are preferentially photometric binaries \citep{Simonian19}, and are
consistent with being a population of tidally-synchronized binaries. 

Rotation can be measured in two different ways. One method measures
the projected broadening of spectral lines due to rotation \citep{Kahler89},
which requires spectroscopy.  An additional method
which has been expanded greatly by the \Kepler{} satellite is the measurement
of rotation periods through modulations in the light curves due to starspots
\citep[see][and references therein]{McQuillan14,Garcia14, Aigrain15,Luri18}.
These measurements take advantage of long stretches of precise and continuous
coverage. Rotational periods are sensitive to
much slower rotation rates than \vsini{} because rotational broadening becomes
comparable to the few~\kms{} resolution of most high-resolution spectrographs. 
However, periods can only be measured for cool stars with spots, and detections at
low levels become challenging because of systematic uncertainties in the
differential photometry (see \citealt{vanSaders19} for a discussion)

There is also little known about
how the behavior of stellar rotation changes with metallicity. Angular momentum evolution models are largely based on open clusters with metallicities in the range of \(-0.01 \le \feh \le 0.16\) \citep{Netopil16}.
Multiplexed high-resolution spectroscopy, such as that done by the Apache Point Galactic Evolution Experiment (APOGEE) will
be useful for measuring metallicity and \vsini{} simultaneously for large
samples.

While there have been several attempts to validate the rotation periods of
\Kepler{} stars by cross-comparisons of different periodogram analyses
\citep[see][]{Aigrain15}, there
have been few efforts to do so with independent measures
such as \vsini{} \citep[see][for a comparison in M dwarfs]{Gilhool18}. The advent of large spectroscopic
surveys, such as the APOGEE, has
enabled the measurement of \vsini{} for hundreds of thousands of stars.

In this paper, we compare the rotation distribution of \Kepler{} targets to
that seen from APOGEE \vsini{}. Section~\ref{sec:methods} describes the sample
selection as well as the methodology used in this analysis. 
Section~\ref{sec:validation} describes the validation done for the APOGEE
\vsini.  Section~\ref{sec:results} presents the main science results for the
\Kepler{} subgiants. We then discuss the implications of the study and conclude
in Section~\ref{sec:discussion}. 

\section{Methods}
\label{sec:methods}

In this work, we compare APOGEE \vsini{} to \Kepler{} rotation periods and to models
of stellar rotation.
In order to perform these comparisons, we supplement rotation data with stellar
properties, such as temperature, stellar radius, metallicity, and luminosity.
We begin by describing the sources of data for our primary analysis, the sample selection for our main science sample, and finally describing methods and additional datasets for data validation..

\subsection{Fundamental Stellar Properties}
\label{sec:stelprops}

The stellar parameters we are interested in are temperature, chemical
abundance, luminosity, and radius. While some are measured directly from
the spectra, others need to be inferred indirectly. We describe how these parameters
are derived in detail below.

\subsubsection{Effective Temperature, Metallicity, and Alpha Enhancement}

The underlying sample is drawn from Data Release 14 \citep{Abolfathi18}
of the Sloan Digital Sky Survey (SDSS) \citep{Majewski17}, hereafter DR14. Our
data is obtained with the APOGEE spectrograph, a critically-sampled 
high-resolution (\(R \sim 22,000\)) multi-fiber near-infrared (1.51--\(1.70 \mu m\)) spectrograph 
\citep{Wilson10} mounted on the SDSS 2.5-meter telescope at Apache Point 
Observatory \citep{Gunn06}. The APOGEE Stellar Parameters and Chemical 
Abundances Pipeline (ASPCAP) measures stellar parameters, such as \Teff{}, \feh{}, and \afe{}, as well as rotational
broadenings by fitting the APOGEE spectra to a grid of stellar atmosphere models.

\citet{Serenelli17} compared APOGEE DR13 \Teff{} and \feh{} values 
for the APOGEE-\Kepler{} Asteroseismic Science Consortium (APOKASC) dwarfs and subgiants to 
those measured from high-resolution optical spectroscopy
and found that the temperatures were consistent within the uncertainties of
around 80 K, with a possible 20 K zero-point shift, which is very small for our
purposes. However, they claimed that the reported uncertainty in \feh{} was
underestimated, and the true random error was closer to 0.1 dex. No analogous
analysis was performed in DR14, and we therefore adopt the
\citet{Serenelli17} errors.

\subsubsection{Binarity}
\label{sec:binarity}

\begin{figure*}[htb]
    \plotone{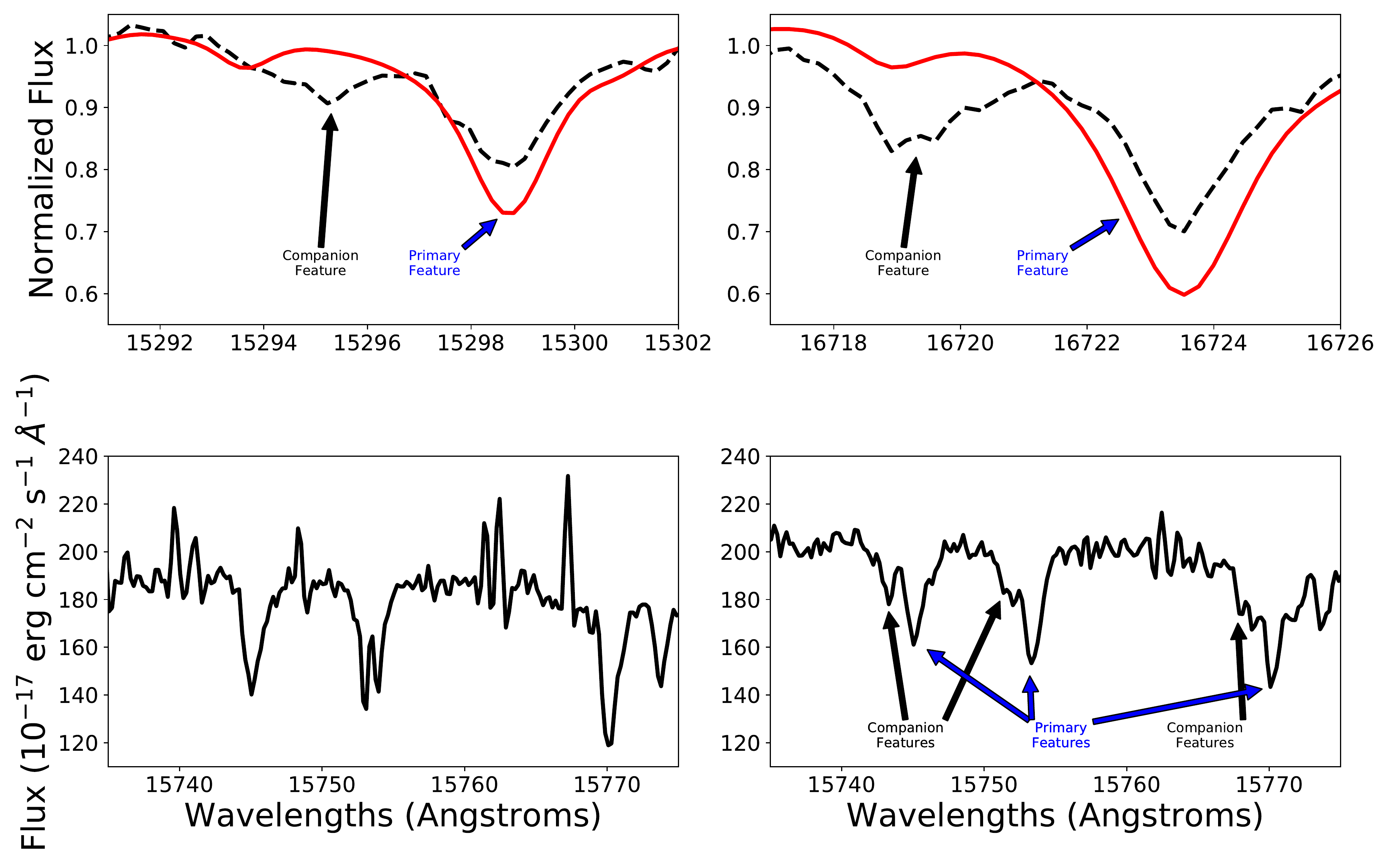}
    \caption{Illustrative examples of visual SB2s. \emph{Top Left:} Combined 
        spectrum for KIC 11856076 centered on 15296.5~\AA{}. Two features are 
        separated by 3.5~\AA{}. The combined APOGEE spectrum is shown as a dashed black line, while the best-fit ASPCAP spectrum is shown as
a solid red line. \emph{Top Right:} Same spectrum as top left except centered on
16721.5~\AA{}. There are also two features which are separated by approximately 
4~\AA{}.  \emph{Bottom Left:} Visit spectrum for KIC 7663322 in region centered 
on 15755~\AA{} containing many features. \emph{Bottom Right:} Visit spectrum 
taken 538~days later with second component visible.}
    \label{fig:sb2diag}
\end{figure*}

Most stars are found in binary systems \citep{Moe17}, and main sequence 
binaries have been shown under some circumstances to yield biased stellar 
parameters when analyzed as single stars \citep{ElBadry18a,Simonian19}. By 
contrast, the steeply mass-dependent main-sequence lifetime means that
evolved primaries in binaries are more luminous than their companions, so the
impact of binarity is reduced. In our sample, we can compare the impact of
binarity on stellar parameters of evolved stars compared to those near the main sequence. We use two main methods 
to flag binaries in our
sample: visual inspection of the spectra for SB2s, and inference of photometric binaries from excess luminosity above the lower main sequence.

\textbf{Double-Lined Spectroscopic Binaries (SB2)}: One potentially significant class of contaminants are SB2s. The ASPCAP pipeline assumes a single-star
stellar atmosphere to derive stellar parameters. If a spectrum has two sets of lines, 
ASPCAP will attempt to fit an extremely broad single line profile on top of both sets 
of lines, resulting in a spurious rapid rotation measurement. We visually inspected the 
spectra of all targets in our sample with 
\vsini{} greater than \(10~\kms\) (the detection threshold determined in
Section~\ref{sec:pleiades}), and flagged them if the star showed 
signatures of double-lined spectra
either in the combined spectrum, or in any of the individual visit spectra. 
An illustrative example of each case is shown in \cref{fig:sb2diag}. As expected, we find evidence
that the SB2s have overestimated \vsini{} compared to rotation periods
in Section~\ref{sec:cooldwarfs}.

\textbf{Photometric Binaries}: We separate photometric binaries from single stars on the lower main sequence
following the methodology described in \citet{Simonian19}, using the
conservative threshold of 0.3 mag above the main sequence. One
difference in methodology is that we adopt version 1.2 of the MIST isochrones
(described in Section~\ref{sec:radius}) instead of version 1.1.

We impose an upper limit of luminosity at \(-1.3\) mag above the main sequence,
which corresponds to three times the main-sequence luminosity. This ceiling
leaves ample room for excess luminosity in potential triple systems 
\citep{Simonian19}. Targets more luminous than this
ceiling are labeled as ``sub-subgiants'' (see \cref{fig:rawdata}), and they are
discussed in Section~\ref{sec:redstragglers}.

\subsubsection{Luminosity}
\label{sec:luminosity}

We calculate bolometric luminosity by combining \Gaia{} parallaxes
with 2MASS \Ks{} photometry \citep{Skrutskie06} and extinction estimates from 
\citet{Berger18b},
and then adding in a \Ks-band bolometric correction. Anytime we illustrate the sample 
in an HR diagram, we use the de-reddened \Ks-band luminosity as a precise marker 
of evolutionary state, minimizing the impact of extinction.

We also use high-quality parallaxes from the \Gaia{} mission's Data
Release 2 \citep{Gaia18}. We match \Kepler{} targets against \Gaia{} DR2 using
the cross-matched database of \citet{Berger18b}, which contains all but 778 out 
of the 16,915 \Kepler{} targets in DR14. The missing targets can be attributed to 
the quality cuts in \citet{Berger18b}\footnote{\citet{Berger18b} excluded targets with parallax errors greater than 20\%, \(\Teff < 3000\) K, \(\logg > 0.1\) dex, and targets with less than ``AAA``-quality 2MASS photometry}. Because
\citet{Berger18b} only included targets with low parallax uncertainties, we use the traditional formula 
for deriving distance modulus from parallax instead of the Bayesian method 
advocated by \citet{Luri18} to reduce computational complexity. We include a 
zero-point offset of 50.2 \(\mu\)as determined from comparisons to 
asteroseismic radii \citep{Zinn19}.

We use estimates of extinction from the \citet{Berger18b} catalog, which are
based on the 3-D reddening map of
\citet{Green18}. We convert the extinction to \(A_K\) using the 
\citet{Cardelli89} relation that \(A_K/A_V = 0.114\). 

To get the bolometric absolute magnitude, 
we add \MK{} to the bolometric correction calculated by interpolating the ATLAS12/SYNTHE 
\citet{Kurucz70, Kurucz93} stellar atmosphere models included with the MIST
isochrones over \Teff{} and \feh{}. 

A very small fraction of targets (0.3\%) of the 
APOGEE sample have 2MASS photometry flagged due to blending. We exclude these
stars. 

\subsubsection{Radius}
\label{sec:radius}

\textbf{Single Stars:} We first calculate isochrone-independent 
radii by making use of the bolometric luminosity and temperature through the 
Stefan-Boltzmann law. Calculating radii makes use of the bolometric absolute 
magnitude from Section~\ref{sec:luminosity}, and the effective temperature, 
which are available in the combined APOGEE/\Gaia{}/2MASS catalogs. 

\textbf{Photometric Binaries:} The above method will yield incorrect radii 
for unresolved photometric binaries because the Stefan-Boltzmann law assumes all of the flux comes 
from a single star. Excess luminosity from a binary companion will result in the radius of the primary star being overestimated. In general, radii depend on both effective temperature and age, so we cannot easily disentangle evolved subgiants from photometric binaries. However, on the unevolved lower main sequence (\(\Teff < 5250 \textrm{ K}\); see
\citealt{Simonian19}), age effects are small and we can use the main-sequence radius-effective temperature relationship to estimate the true radii. We perform this correction by projecting the
observed \Ks-band absolute magnitude onto a 1 Gyr, \(\feh=0.08\) MIST isochrone 
at fixed \Teff{}, as done in \citet{Simonian19}.

\subsection{The APOGEE-\Kepler{} Dwarf and Subgiant Sample}
\label{sec:sample}

\begin{deluxetable}{l c c}
\tablecaption{APOGEE Targeting Flags\label{tab:targeting}}
\tablehead{\colhead{Targeting Flag} & \colhead{Total} & \colhead{Dwarfs/Subgiants}}
\startdata
\APOGEECOOLDWARF & 702 & 598 \\
\APOKASCDWARF & 2303 & 2265 \\
\APOKASCGIANT & 4114 & 388 \\
\APOGEEKOI & 340 & 329 \\
\APOGEEKOICONTROL & 38 & 36 \\
\APOGEESEISMO & 7509 & 1486 \\
\APOGEERVMONITOR & 163 & 162 \\
\APOGEEHOST & 111 & 108 \\
\hline
\hline
Total & 15242 & 5337
\enddata
\tablecomments{
Number of stars in each APOGEE targeting program as of DR14, and those that pass the Dwarf/subgiant cut. Some stars were targeted in 
multiple programs, so the total sample will be less than the sum 
of all targeting flags.}
\end{deluxetable}

We focus on the sample of \Kepler{} stars observed by APOGEE as of DR14
\citep{Zasowski17}. The majority of the \Kepler-APOGEE
targets were chosen for asteroseismic studies of evolved red giants, but 7,918 
unevolved stars were targeted by the survey, and 5,337 were observed as of DR14. 

\begin{figure}[htb]
  \centering
  \plotone{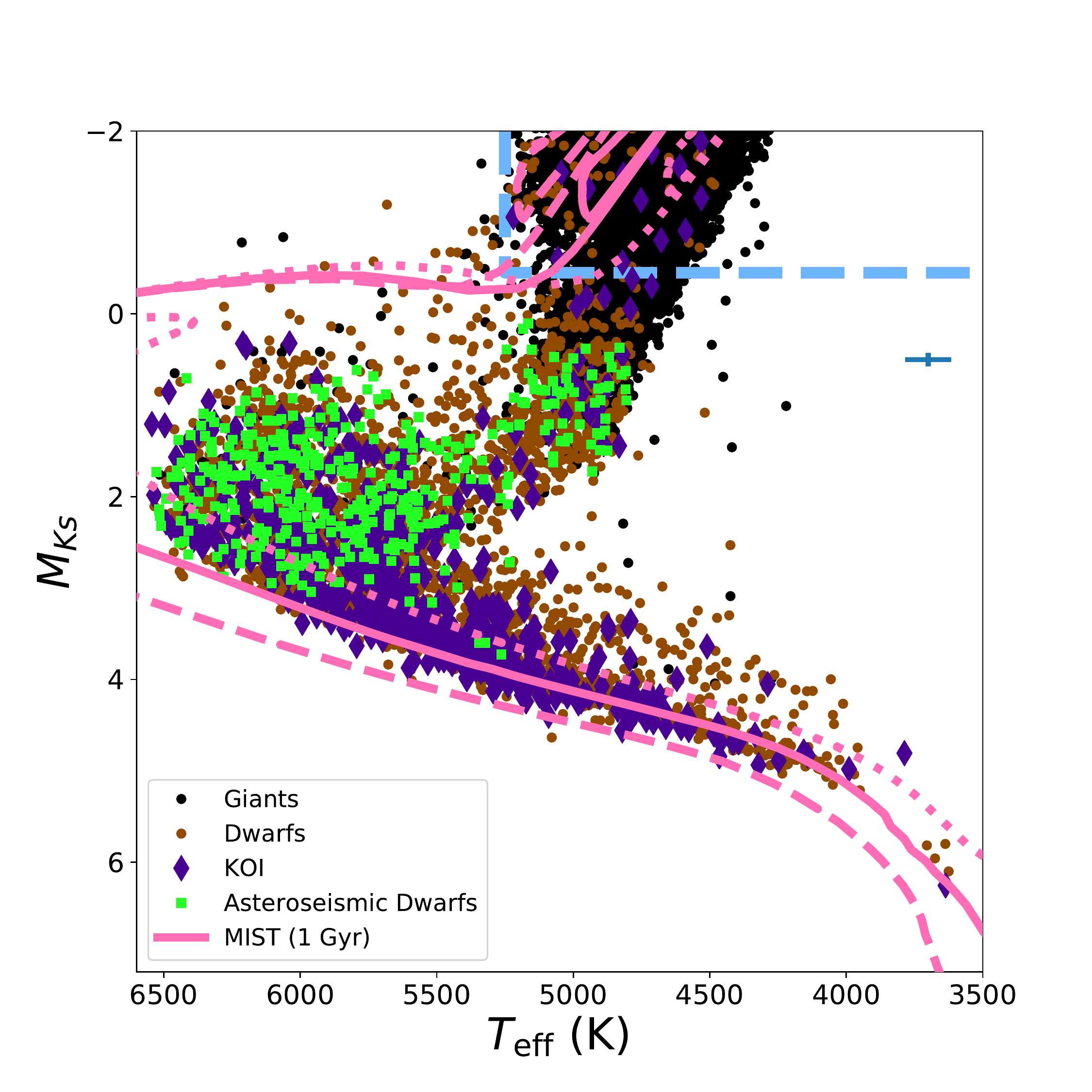}
  \caption{The APOGEE sample based on targeting flags. Stars
      flagged as part of the giant sample as black dots, dwarf sample as red 
      dots, and \Kepler{} Objects of Interest as purple diamonds (see text for
      details). Targets with asteroseismic radii analyzed in 
      \citet{Serenelli17} are marked as green triangles. Our selection cut for
      giants is shown as the blue dashed line. A representative error 
      bar is shown in the top right. 1 Gyr MIST isochrones for metallicities 
      \(\feh = -0.5, 0.0, \textrm{ and } 0.5\) are shown in pink dashed, solid, 
      and dotted lines, respectively.\label{fig:targeting}}
\end{figure}

The majority of our targets were taken from magnitude and color selected 
samples of dwarfs and subgiants, supplemented by a smaller 
asteroseismic data set. We also included several
smaller surveys which targeted dwarfs. For full details about the targeting 
in APOGEE, we refer the reader to the APOGEE targeting 
page\footnote{\url{https://www.sdss.org/dr14/irspec/targets/}}, as well as 
\citet{Zasowski17}. We provide a brief overview of the selection functions here. 

Stars targeted for specific science programs in APOGEE are flagged as such. The programs we draw
from are briefly described here, and listed in \cref{tab:targeting}. The \APOGEECOOLDWARF{} targets are bright (\(H \le 11\) mag), with \citet{Pinsonneault12} \(\Teff < 5250\) K and \citet{Brown11} \(\logg > 4.0\). \APOKASC{} targets extended this sample to hotter temperatures and lower surface gravities based on \citet{Pinsonneault12} \Teff{} and \citet{Huber14} \logg{}. \APOKASC{} is made up of \APOKASCGIANT{}, which selects stars with \(\Teff < 5500\) K and \(\logg < 3.5\), and \APOKASCDWARF{}, which selects stars with \(6000 \textrm{ K } > \Teff > 4500 \textrm{ K }\). \APOGEESEISMO{} targets were selected to follow-up stars showing asteroseismic oscillations in \Kepler{}. Most of this sample is giants, but 426 stars were dwarfs targeted for asteroseismic studies \citep{Chaplin11}. \APOGEEKOI{} and \APOGEEKOICONTROL{} were selected to follow-up \Kepler{} planet host stars. The number of
stars in our sample from each of the targeting programs is shown in \cref{fig:targeting}.

After gathering all targets from \cref{tab:targeting}, we remove giants by
imposing a cut on targets with \(\Teff < 5250\) K and \(\MK < -0.45\), shown in
\cref{fig:targeting}. This cut is more inclusive of the base of the RGB than a traditional
spectroscopic \(\logg > 3.5\) cut. However, we choose to be more inclusive because of the discovery of a population of potential red stragglers (see Section~\ref{sec:redstragglers}). While  \citet{Tayar15} has done a more thorough investigation of the giants which is not our focus, the APOGEE sample at the time of that study was not large enough to reveal this red straggler population. The number of dwarfs and subgiants passing 
this cut are we also listed in \cref{tab:targeting}.

Our final dwarf/subgiant sample has the following properties: \(7 \le H \le 14\). The typical range of parallaxes
(with \(1\sigma\) quartile ranges) is \(1.86^{+2.55}_{-1.02} \) mas. The median extinction 
 for
our sample of dwarfs and subgiants is \(0.013^{+0.014}_{-0.006}\) mag.

\citet{Huber14} added stars to the KIC with saturated KIC photometry by
leveraging 2MASS colors. Because APOGEE used the \citet{Huber14} catalog for
selection, it targeted a substantial fraction of these stars---34\%---which
did not previously have stellar parameter estimates. We evaluate the 
reliability of the \citet{Huber14} methodology in Appendix~\ref{sec:huber}.

\subsubsection{Spectroscopic Flags and Quality Cuts}

The ASPCAP pipeline infers stellar parameters \citep{GarciaPerez16} by fitting 
a 6-dimensional grid of stellar atmospheres to the APOGEE spectra 
\citep{AllendePrieto06}. This fit failed significantly for 2.6\% of our sample 
and triggered
the \STARBAD{} quality flag\footnote{The \STARBAD{} flag is triggered if any of the stellar parameters are near the grid edge, the combined S/N of the spectrum is less than 50, the chi-square of the fit is greater than 50, or the derived spectroscopic temperature is discrepant from the \citet{GonzalezHernandez09} \(J-K\) color temperature by more than 500 K. For more information see the APOGEE documentation at \url{https://www.sdss.org/dr14/irspec/parameters/}.}. Less extreme discrepancies can trigger the
\STARWARN{} flag. Visual inspection of a representative sample of targets with
\STARWARN{} indicated reasonable fits, so we included these targets and
excluded almost all objects with the \STARBAD{} flag.

However, there are 13 objects in our sample where the \STARBAD{} flag was only 
triggered because the \vsini{} hit an upper grid edge in the fit. To avoid
preferentially excluding rapid rotators, we manually calibrate stellar 
parameters for these stars \citep{Holtzman18}, and treat those few targets as 
\vsini{} lower limits. This sample is so small that inclusion of these stars do not substantially affect our conclusions.

\subsection{Validation Datasets}
\label{sec:valdata}

\begin{deluxetable*}{l c c}
\tablecaption{Validation Sample Overlap\label{tab:validation}}
\tablehead{\colhead{Sample} & \colhead{Total Size} & \colhead{Overlap with APOGEE}}
\startdata
\citet{McQuillan14} Detections & 34030 & 964 \\
\citet{McQuillan14} Nondetections & 99000 & 1476 \\
\citet{Serenelli17} & 426 & 393 \\
\citet{Garcia14} & 310 & 220 \\
\citet{ElBadry18b} & 20142 & 904 \\
\citet{Berger18a} & 177911 & 5337
\enddata
\tablecomments{
Overlap between validation samples and our main APOGEE sample
described in Section~\ref{sec:sample}.}
\end{deluxetable*}

Some of the data used in Section~\ref{sec:stelprops} is relatively new and has not been extensively validated, we perform additional validation by supplementing these data with that from other studies. Comparison to rotation periods is a major part of our validation, but we also supplement with other datasets of radius, binarity and temperature. For a summary of the validation tables, and the size of their overlap with the APOGEE rotation sample, see \cref{tab:validation}.

\subsubsection{Rotation}

The APOGEE \vsini{} measurements were only included as 
catalog values as of DR14.  As a result, they have not been thoroughly
validated. \citet{Tayar15} measured \vsini{} directly from the APOGEE spectra 
and showed that rapid rotation could be successfully measured in giants. We
extend this validation into the dwarf and subgiant regime by comparing to 
rotation periods, as well as to literature \vsini{} (see
Section~\ref{sec:pleiades}). We describe how we compare the \vsini{} and
rotation period in this section.

The main sample of rotation periods in this work comes from
\citet{McQuillan14}. \citet{McQuillan14} selected their initial sample using
the \Teff-\logg{} cuts advocated by \citet{Ciardi11} to distinguish between 
dwarfs and giants. In order to avoid 
confusion with other periodic signals, \citet{McQuillan14} removed known 
eclipsing binaries \citep{Prsa11,Slawson11}, KOIs \citep{Akeson13}, and
imposed a maximum \Teff{} of 6500 K to reduce confusion with 
pulsators. Their sample to be analyzed contained 133,030 stars.

\citet{McQuillan14} calculated rotation periods using an autocorrelation
function (ACF) based method. The result 
was a list of 34,030 stars with period detections and 99,000 stars without
them.

In addition to the \citet{McQuillan14} periods, we also make use of surface 
rotation periods derived for a sample of 310 \Kepler{} stars showing
asteroseismic oscillations \citep{Garcia14}, 217 of which have been observed
with APOGEE\@. The \citet{Garcia14} periods provide a larger overlap with the sample
of stars with asteroseismically-derived radii \citep{Serenelli17} than \citet{McQuillan14}.

\subsubsection{Transformation to \(v_{eq}\)}
\label{sec:transform}

\vsini{} and rotation period are not directly comparable quantities. However,
they can be compared by transforming rotation periods to equatorial velocities
through the equation:
\begin{equation}
    \log v_{eq} = \log 2 \pi R - \log P
    \label{eq:veq}
\end{equation}
where \(R\) is the stellar radius, and \(P\) is the rotation period.

\begin{figure}[htb]
    \centering
    \plotone{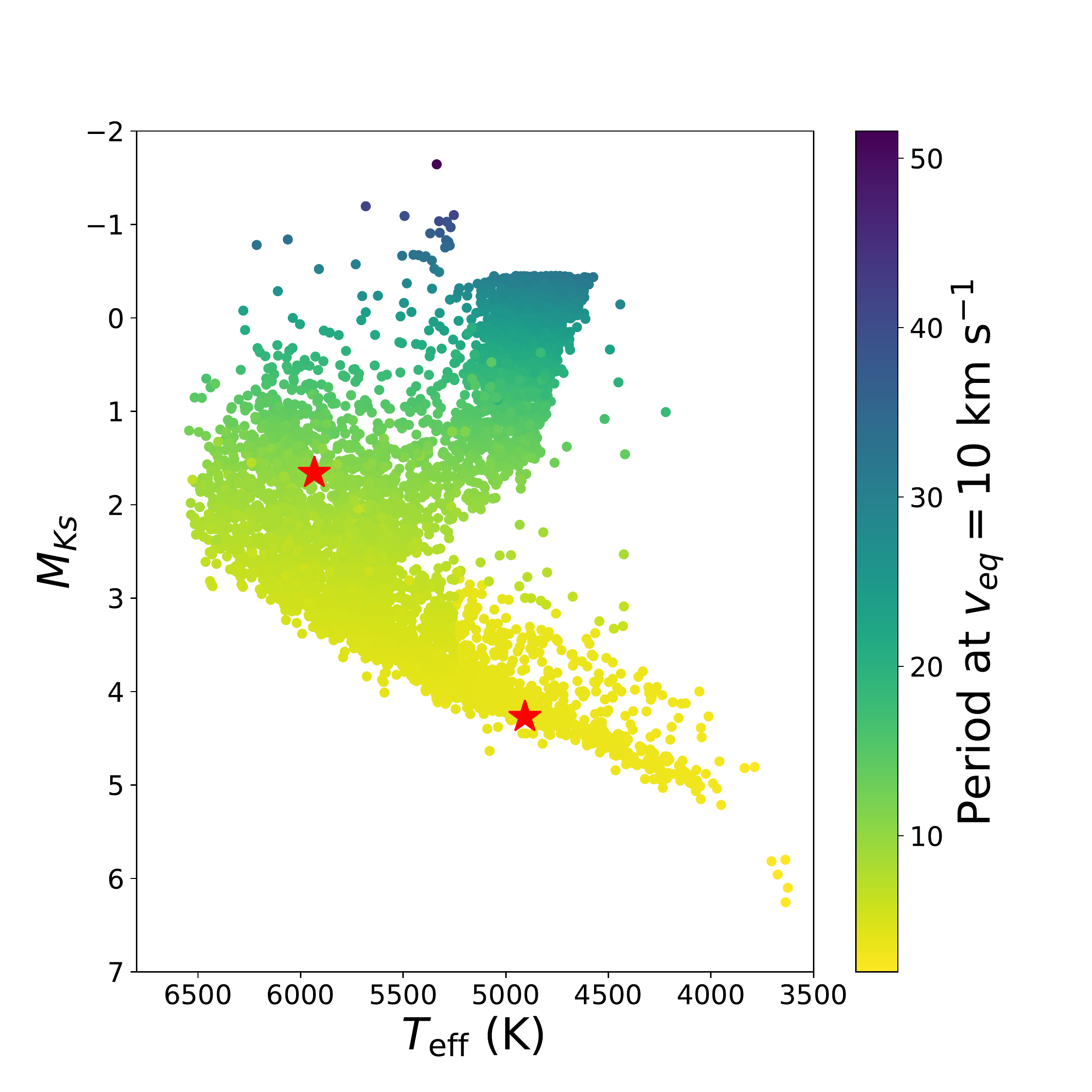}
    \caption{HR diagram showing how a detection threshold in velocity
        translates to period as a function of HR diagram position. The
        Stefan-Boltzmann radius is used to translate the velocity to a period, 
        except for the cool dwarfs where the deprojected radius is used.
        Red stars denote the median period threshold in the cool dwarf sample
        (3.7~days), and in the asteroseismic sample 
    (6.7~days).\label{fig:period_vel}}
\end{figure}

This transformation using the stellar radius means that a flat detection limit 
of 10~\kms{} (found in Section~\ref{sec:pleiades}) translates to a 
radius-dependent period threshold illustrated in \cref{fig:period_vel}. 
The period cutoff is 3.7~days for a \(R = 0.74~R_\sun\) star, typical of the 
cool dwarf sample; and 6.7~days for a \(R = 1.3~R_\sun\) star,
typical of the asteroseismic sample. 

The \(\sin i\) ambiguity means that \vsini{} and period are best compared in a
statistical sense for large samples. For individual stars, a \vsini{} greater
than \(v_{eq}\) implies \(\sin i > 1\), which is unphysical. However, individual values
with \(\sin i \le 1\)  cannot be directly ruled out.

After calculating the equatorial velocity given the period
and radius, we construct a distribution by convolving it with a \(\sin i\) 
distribution, and calculate how much of the convolved distribution remains above the
\vsini{} detection limit of 10~\kms. Both period detections and
\vsini{} are biased against low \(\sin i\): spot modulation cannot be seen for
low inclination and \vsini{} drops below the instrumental resolution or other
broadening mechanisms when the inclination is low enough. Based on prior 
comparisons in star clusters we truncate our \(\sin i\) convolution kernel at 
\(\sin i = 0.5\) \citep{Krishnamurthi97,Jackson10} to correct for this shared
bias. We can then compare the observed spectroscopic rapid rotator fraction to that predicted
by the photometric rapid rotator fraction in selected populations to serve as an external 
consistency check between the two measures of rotation. We use this average property, as opposed to a star by
star measurement, because it incorporates the known statistical properties of 
\(\sin i\). When calculating the spectroscopic rapid rotator fraction we do not include
visual SB2s, their \vsini{} will be spuriously overestimated. 

\subsubsection{Binarity}

To evaluate our method of identifying binarity in Section~\ref{sec:binarity}, 
we compare the SB2s derived this way to those determined by \citet{ElBadry18b}. 
\citet{ElBadry18b} ran a machine-learning algorithm to flag
stars with spectra inconsistent with a
single star model.  For stars with poor single-star fits, \citet{ElBadry18b} 
attempted to find a better-fitting model including a binary companion. 
\citet{ElBadry18b} ran their analysis on the APOGEE DR13 release, which 
only contained 17\% of the \Kepler{} dwarfs and subgiants in DR14,
necessitating the visual inspection previously discussed. 

When tested in the overlap sample of 308 targets, the \citet{ElBadry18b} 
methodology flags visual SB2s exceedingly well.
Of the 36 overlap stars visually classified as SB2s, none were classified by 
\citet{ElBadry18b} as single stars, suggesting that the \citet{ElBadry18b}
algorithm is complete. In the remaining sample, \citet{ElBadry18b}
flagged an additional 60 stars as SB2s, which is consistent with their method 
being more sensitive to fainter secondary stars than a visual inspection. 

In addition to flagging binaries, \citet{ElBadry18b} also provide separate 
stellar parameter estimates for the primary and secondary. We use these stellar 
parameters in Section~\ref{sec:cooldwarfs} to illustrate the bias imposed by 
stellar companions on APOGEE stellar parameters, which can be significant.

The results of the comparison show both that the methodology of \citet{ElBadry18b} is extremely reliable at detecting double-lined spectroscopic binaries, and that the visual SB2 identification does not detect a substantial population of false positives. Unfortunately, because DR14 added many new observations of dwarfs and subgiants which were not in DR13,  the \citet{ElBadry18b} sample is not useful for flagging the majority of SB2s in our sample. Without a comprehensive, reliable method of flagging SB2s, we fall back on identifying SB2s by eye, and assume that the sample of SB2s which are detectable by \citet{ElBadry18b} but without visible distortion to the spectra will not have dramatically biased stellar parameters.

\subsubsection{Temperatures}

In comparisons using the full \citet{McQuillan14} sample, most of which do not 
have APOGEE spectroscopic temperatures, we use the
\citet{Berger18b} photometric temperatures, which are available for nearly the
entire \Kepler{} sample, and are based on the fundamental infrared flux method,
the same system as that used for APOGEE.

\section{The APOGEE \vsini{} sample}
\label{sec:validation}

\begin{figure}[htb]
    \centering
    \plotone{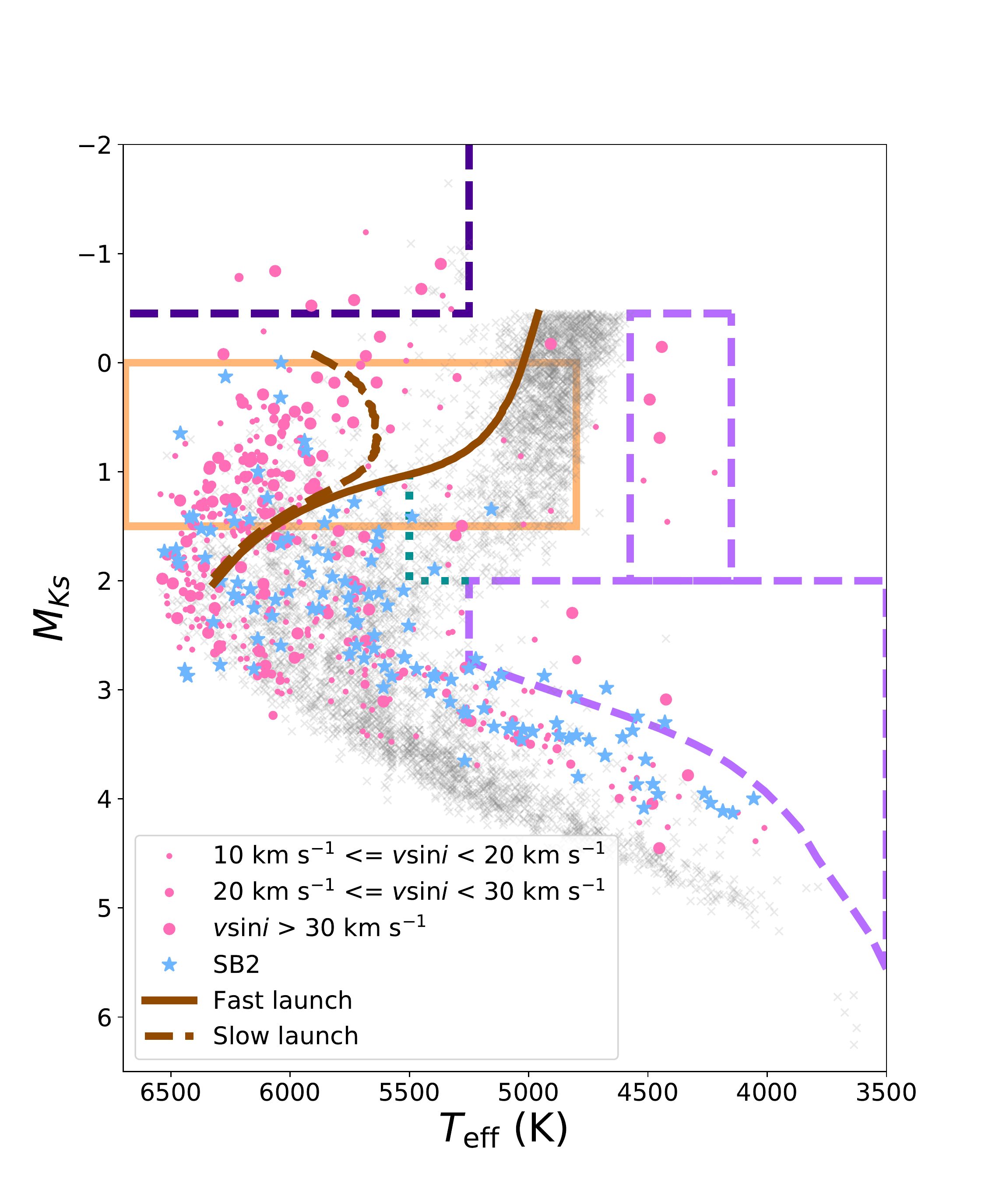}
    \caption{Placement of high \vsini{} objects on the HR diagram. The sizes of
    the marker correspond to bins of \vsini{}. Stars flagged as
visual SB2s are light blue stars. Slow rotators are marked as grey crosses.
Overplotted are the solar metallicity boundaries predicted by the models in
\citet{vanSaders13} for the fast and slow launch conditions (solid and dashed
brown lines), respectively. These launch conditions represent the initial rotation periods and disk-locking timescales required to match the slowly rotating and rapidly rotating sequences of the Pleiades and M37, respectively. This boundary is quantitatively studied within the
orange box in Section~\ref{sec:subgiants}. Also highlighted are blue stragglers 
(within dashed purple lines), red stragglers and sub-subgiants (within violet 
lines). The portion of the sample not accessible by the \citet{vanSaders13} 
models is split into regions with high and low rapid rotator fractions (denoted 
    by the aquamarine dotted line).}
    \label{fig:rawdata}
\end{figure}

\begin{figure*}[htb]
    \centering
    \plottwo{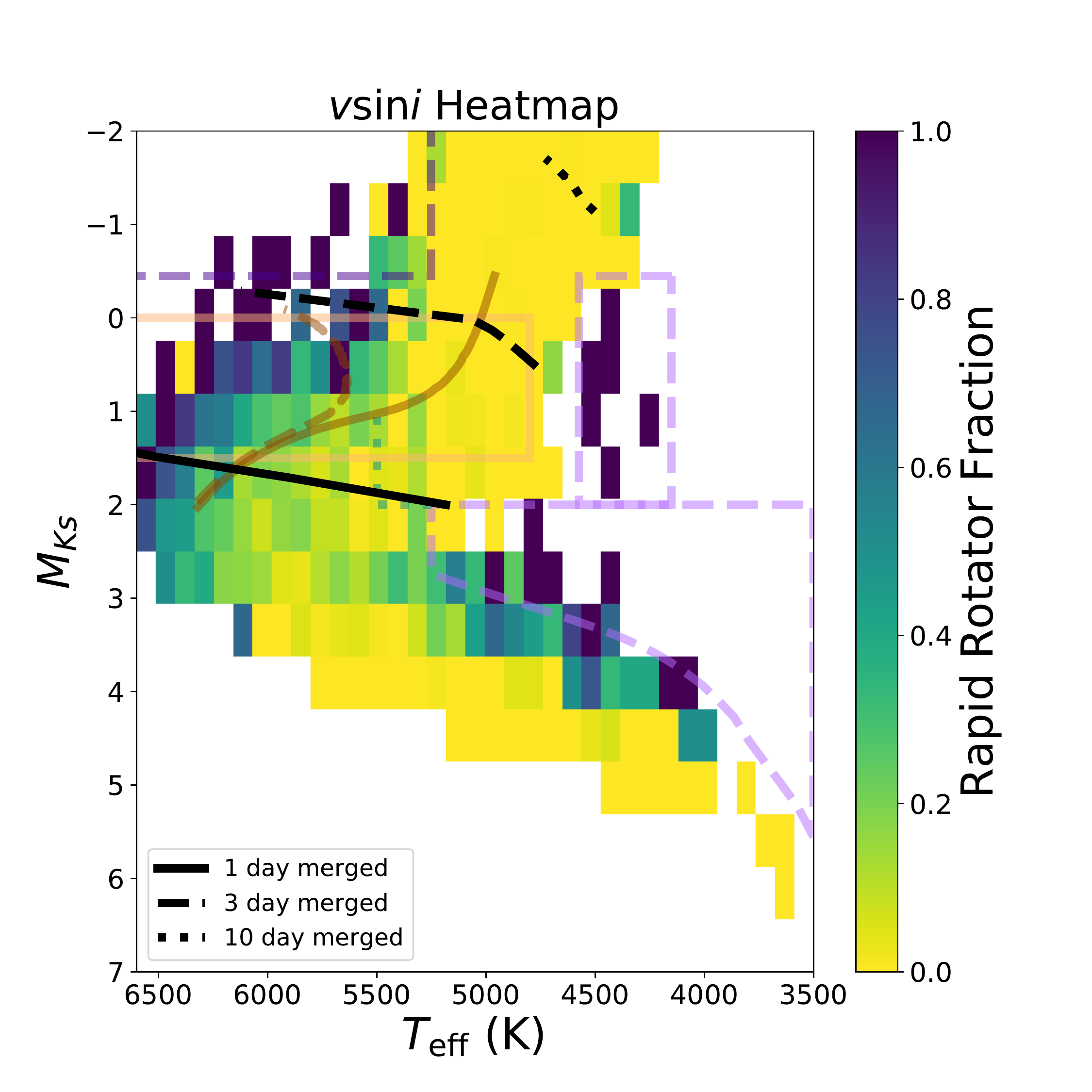}{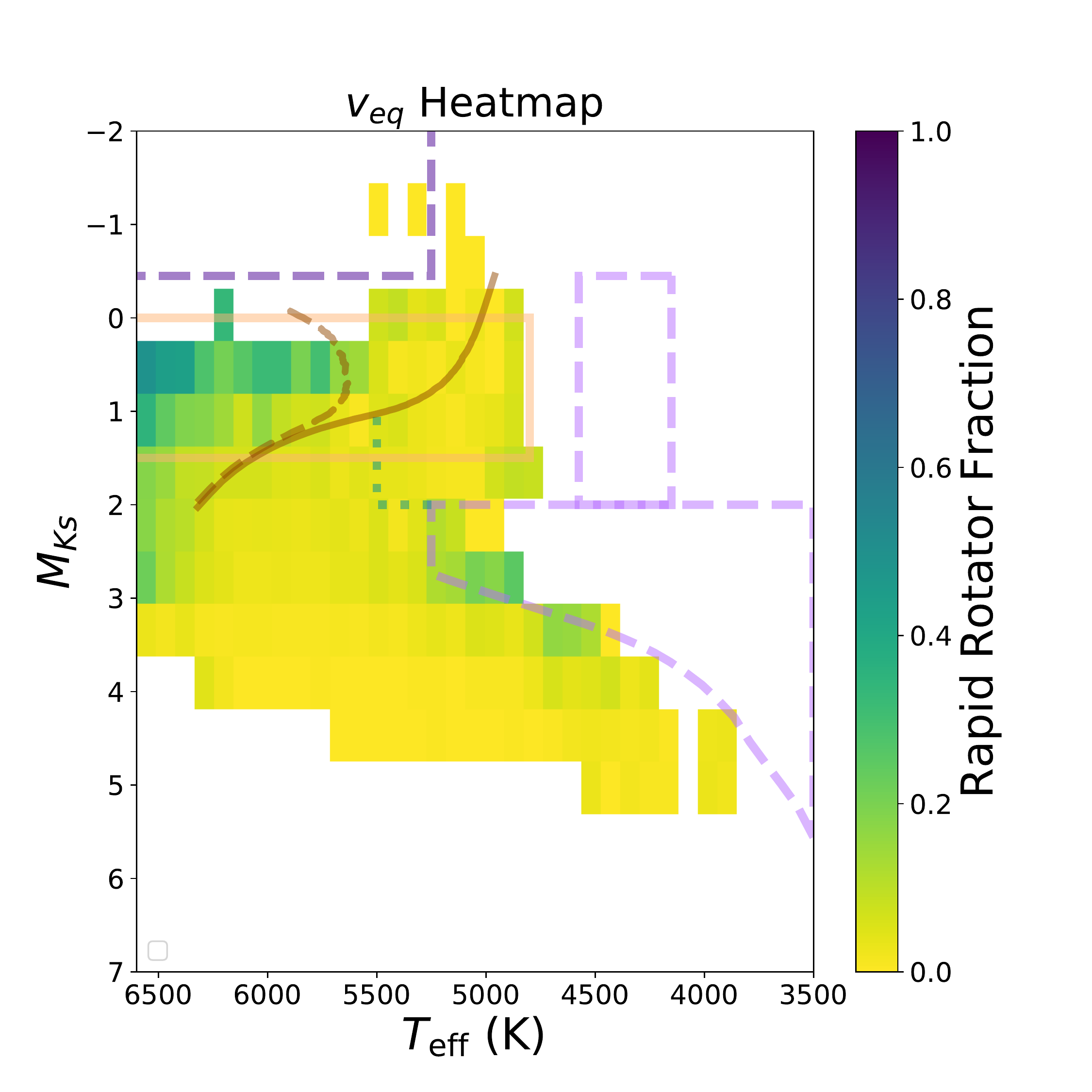}
    \caption{\emph{Left:} Map of the rapid rotator fraction in the HR diagram
        for the APOGEE sample using a threshold of \(\vsini = 10~\kms\).
        The boundaries from \cref{fig:rawdata} are also reproduced. Black 
        lines mark locations on the HR diagram where binaries on circular 
        orbits with orbital periods of 1 (solid), 3 (dashed), and 10 (dotted) 
        days would undergo Roche-lobe overflow. \emph{Right:} Map of 
        photometric rapid rotator fraction for the full \citet{McQuillan14} 
        sample with a \(v_{eq} = 10~\kms\) threshold transformed to a period
        cutoff based on the HR diagram position. Because of the much larger 
        size of the \citet{McQuillan14} sample, only bins 
        with more than 10 objects are shown. When calculating the rapid rotator 
        fraction, targets which did not have a significant period detection 
    were counted as inactive, slow rotators.  \citet{vanSaders13} model
predictions are the same as on the left.}
    \label{fig:rapidmaps}
\end{figure*}

Our APOGEE sample is illustrated in \cref{fig:rawdata} (individual rapid
rotators highlighted) and \cref{fig:rapidmaps} (rapid rotator fractions
indicated). There are striking trends in the distribution of rapid rotators,
which can be traced to the influence of two distinct populations.
As noted in \citet{Simonian19}, there is a significant rapid rotator background
on the photometric binary sequence consistent with a synchronized binary 
population. Because the \Kepler{} fields are old, there is an absence of young,
single, rapid rotators. There is also a substantial 
population of spectroscopic rapid rotators among hotter stars, reflecting the
Kraft break between rapid rotation in massive stars and slow rotation in low
mass stars. Stars slow down from angular momentum conservation as they cross
the subgiant branch, but magnetized winds can be a significant braking
mechanism as well. The region 
where rapid rotation in subgiants is expected from spindown models (brown; 
described in Section~\ref{sec:subgiants}) is also plotted, in the sense that 
rapid rotation for a reasonable range of main sequence rotation rates should be
confined to the left side of the indicated boundaries. The correspondence
between the two is encouraging. We also note the presence of rapid rotation 
among the red stragglers (stars to the right of the red giant Hayashi track)
and the sub-subgiants (stars between the unevolved lower main sequence and the
subgiants).
These objects appear in star clusters, typically as highly active short period
synchronized binaries about to merge \citep{Geller17,Leiner17}.

In addition to raw counts, the rapid rotator fraction is an interesting
population diagnostic. To define the rapid rotator fraction, we bin the HR
diagram by luminosity and temperature, and calculate the fraction of targets
with \(\vsini > 10~\kms\) in each bin. The map of rapid rotator fractions is
shown in the left panel of \cref{fig:rapidmaps}

We complement the map of spectroscopic rapid rotator fraction with a map of photometric rapid rotator fraction
using the full rotation period sample analyzed by \citet{McQuillan14} in the 
right panel of \cref{fig:rapidmaps}. As we lack APOGEE spectra for most of the
\citet{McQuillan14} targets,
we used the photometric temperatures from \citet{Berger18b}, and the same
procedure for converting HR diagram position into radius and \(v_{eq}\) as for
our main sample. In this diagram we plot the detected rapid
rotator fraction, which is the number of stars rotating more rapidly than the
radius-dependent threshold corresponding to \(v_{eq} > 10~\kms\) (see
Section~\ref{sec:transform}) divided by the total number of objects in the
bin analyzed by \citet{McQuillan14} whether or not a period was detected. Under
the assumption that stars with period nondetections are slow and inactive,
this number approximates the true rapid rotator fraction. For hot stars even 
rapid rotation can produce few spots, so we expect a smaller detection fraction
for them.

Through comparing the rapid rotator fraction in \citet{McQuillan14}, we find
that in both the subgiant and photometric binary populations the spectroscopic 
rapid rotator fraction is higher than the photometric rapid rotator fraction.
This difference can be readily explained by the fact that rapidly rotating stars 
with radiative envelopes should not produced spots. However,
the difference in the cool rapid rotators is troubling because the 
\citet{McQuillan14} sample should be very complete there. We explore the origin 
of this difference in the sample selection section below.

\subsection{Sample Validation}

From prior work \citep{Tayar15} we know that the APOGEE can reliably measure
\vsini{} for evolved stars, and the pattern that we see in the hot subgiants in
our spectroscopic sample appears reasonable. The photometric rapid rotator
fraction is also in accord with that derived from independent tests on the lower 
main sequence \citep{Simonian19}. However, as seen in \cref{fig:rapidmaps}, our 
spectroscopic rapid rotator fraction is implausibly large relative to the 
photometric one for the lower main sequence because one would expect the \(\sin
i\) ambiguity to decrease the spectroscopic rapid rotation fraction relative to
the photometric rapid rotator fraction rather than the other way around. This 
indicates a possible background signal that is erroneously
interpreted as a \vsini{} in some cases. We therefore employ a series of
independent tests of the APOGEE \vsini{} values to understand the source of
this background:

\begin{enumerate}
    \item \textbf{Asteroseismic subgiants} We have precise radii from
        asteroseismic subgiants, making them ideal for testing the concordance
        between \vsini{} and rotation period in the transition zone between the
        giant branch and the upper main sequence.
    \item \textbf{The Pleiades} There is extensive APOGEE spectroscopy in the
        Pleiades star cluster, where there are both numerous rapid rotators and
        large independent data sets. This tests the APOGEE data across the FGK
        main sequence domain.
    \item \textbf{Rotation Periods on the Lower Main Sequence} We can compare
        the rotation periods and \vsini{} values directly in the lower main
        sequence, where we can easily distinguish photometric binaries from
        single stars without the confounding factor of age effects.
\end{enumerate}

As described in our validation below, we conclude that the APOGEE
\vsini{} reliably trace rapid rotation in the subgiant domain, and the
discrepancy in the rapid rotator fractions there is caused by reduced spot 
visibility for hot stars. However, we also find that 13\% of photometric binaries in the \Kepler{} field
have \vsini{} and rotation period that
disagree by at least \(3\sigma\). We 
propose that these disagreements have two sources: one from radius inflation of 
K and M dwarfs, and the other being the presence of unresolved 
spectrally-blended binaries. Because of these
complex systematics, we do not attempt to infer astrophysical properties from
samples near the photometric binary sequence, but restrict our
main science results in Section~\ref{sec:results} to the subgiants.

\subsubsection{Subgiants with Asteroseismic Data}
\label{sec:rotcomp}

We begin our validation using \Kepler{} targets with both asteroseismic radii
and rotation periods. We adopt a 10~\kms{} threshold for calling stars rapid
rotators, derived from the Pleiades test below in Section~\ref{sec:pleiades}.

For the asteroseismic sample, we use rotation periods from \citet{Garcia14}
rather than \citet{McQuillan14} because of the greater overlap. Comparisons
using the overlap sample between \citet{McQuillan14} and \citet{Garcia14} show
that the results are consistent to within their errors, and repeating the
following analysis using the smaller overlap sample with \citet{McQuillan14}
yields similar conclusions.

\begin{figure}[htb]
    \centering
    \plotone{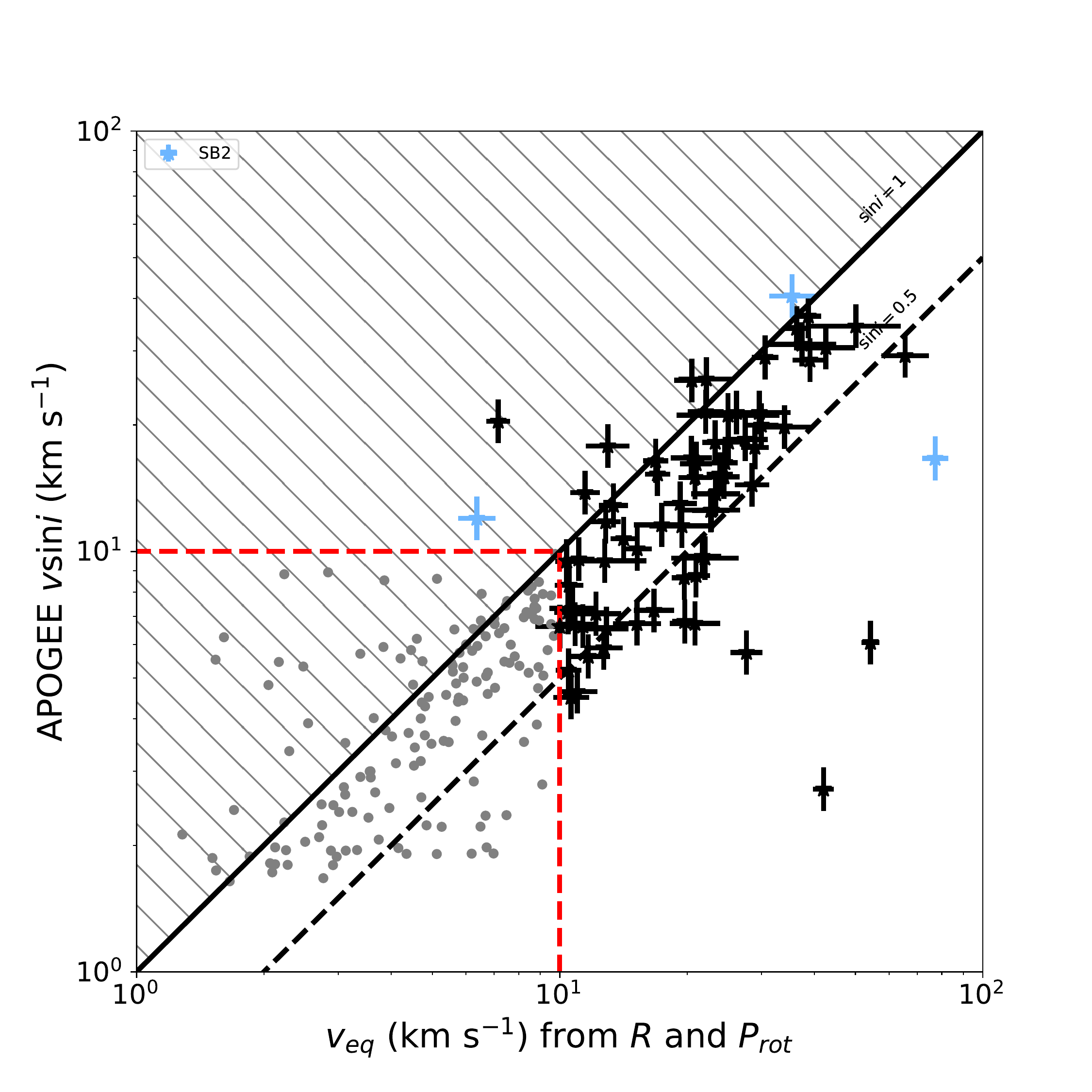}
    \caption{Measured \vsini{} compared to the equatorial velocity
        predicted from rotation period and stellar radius for the asteroseismic
    sample with \citet{Garcia14} rotation periods. Targets above either the
    10~\kms{} \vsini{} or \(v_{eq}\) detection thresholds (shown as dashed red 
    lines) are denoted as bold points. Targets below both thresholds are
    de-emphasized for clarity. Not shown are the 21 stars classified by \citet{Serenelli17} as dwarfs, but with spectra classified by ASPCAP as giant spectra. Blue stars correspond to visually-classified 
    SB2s. The black solid and dashed lines correspond to \(\sin i = 1\) and 
    \(\sin i = 0.5\), respectively. The hatched region corresponds to 
    \(\sin i > 1\), which represents an unphysical region.}
    \label{fig:apokasc_vdists}
\end{figure}

Of the 241 asteroseismic dwarfs with \citet{Garcia14} rotation 
periods, 75 have either \(\vsini > 10~\kms\) or projected equatorial velocities 
greater than \(10~\kms\). There were also 21 stars which were classified by \citet{Serenelli17} as dwarfs, but for which a \vsini{} wasn't calculated by ASPCAP because the spectrum was run with the giant grid. 19 of these stars have equatorial velocities below APOGEE's detection limit, while one has an equatorial velocity of \(12~\kms\). The final star (KIC 11558953) has an equatorial velocity of \(144 \kms\), yet no visual signature of broadening in the APOGEE spectrum. This target has a rotation period of 1.01 days, which is on the edge of the 1--100 day grid which \citet{Garcia14} performed their period search, indicating that this may be a spurious rotation period. Because the overwhelming majority of targets in the \citet{Garcia14} overlap sample without \vsini{} are expected to have non-detectable rotation, we classify them all as slow rotators.

The correspondence between the expected and observed 
velocities for the objects 
with \vsini{} detections is shown in \cref{fig:apokasc_vdists}. The agreement
is generally good, with the single notable outlier (KIC 9908400) being a
clearly blended source in high-resolution imaging; the rotation signal may not
be from the spectroscopic target. There are also three asteroseismic stars (KIC 8656341, KIC 10794845, KIC 10081026) with very low \(\sin i\), where even accounting for APOGEE's \vsini{} detection threshold \(\sin i < 0.5\). KIC 10794845 has clear contamination in the pixel, possibly explaining the two sets of \vsini{} measurements. The other two may be erroneous due to aliasing.

The spectroscopic rapid rotator fraction for the asteroseismic sample is 
\(19 \pm 3\)\% with a predicted spectroscopic rapid rotator fraction 
of \(26 \pm 3\)\% after correcting for inclination effects, 
which are consistent just below the \(2\sigma\) level. If we remove our assumption that \(\sin i > 0.5\), this discrepancy drops to \(1 \sigma\), which leads us to conclude from the asteroseismic subgiant sample that rotation characteristics of a large sample are robust, even though there may be a few individual outliers.

\subsubsection{Pleiades Dwarfs with \vsini{}}
\label{sec:pleiades}

While an ideal validation sample would have rapidly-rotating stars 
representative of the \Kepler{} dwarfs, there are not other field star
spectroscopic data sets with high enough overlap with APOGEE to be useful. 
However, the APOGEE survey obtained a large number of spectra in the Pleiades 
star cluster, which has been the subject of extensive rotational velocity
surveys \citep{Stauffer87,Queloz98,Jackson18}. Because of its youth, rapid 
rotation is a general phenomenon there. There is also a large data set of literature
\vsini{} available, which we adopt to test the APOGEE \vsini{} values for 
active dwarfs. 

\begin{deluxetable}{ l c c }
\tablecaption{APOGEE Overlap with Previous Literature\label{tab:pleiadescount}}
\tablehead{\colhead{Survey} & \colhead{R} & \colhead{Overlap}}
\startdata
\citet{Queloz98} & 30000 & 91 \\
\citet{Stauffer87} & 40000 & 63 \\
\citet{Jackson18} & 17000 & 91 \\
TOTAL &  & 144
\enddata
\tablecomments{Total number is much smaller than the sum of surveys because of large overlap between surveys. For surveys which used multiple instruments, the lowest resolution in the study is displayed.}
\end{deluxetable}

We select stars which were observed both in APOGEE as well as in 
at least one of these previous studies, cross-matching using the master table 
in \citet{Rebull16a}, which yields a total overlap of 245 observations of
144 targets. A detailed breakdown of the overlap with each survey is given in 
\cref{tab:pleiadescount}, and the cross-matched sample is given in \cref{tab:pleiadessample}. 

To better characterize the Pleiades sample, we supplement the rotation data
with 2MASS photometry \citep{Skrutskie06}, and look up \Gaia{} parallaxes by 
cross-matching the 2MASS IDs in \citet{Rebull16a} with the \Gaia{} external 
cross-matched catalog \citep{Marrese19}. All of the targets in 
\citet{Rebull16a} that were also observed in APOGEE have parallaxes in \Gaia{}.

In an initial comparison, we found that the agreement with the literature was
excellent for single stars, but that there were some anomalies for binaries. We
therefore separate the sample into photometric single and binaries stars, and
illustrate their properties separately. We consider objects in the Pleiades to 
be photometric binaries if they are more than 0.3 mag above the
main-sequence locus in \(\Delta K\) vs \(V-K\) as measured by 
\citet{Stauffer16}. 31 out of the 144 targets in this sample (or 22\%) are
photometric binaries.

\begin{figure*}[htb]
  \centering
  \includegraphics[width=1.0\textwidth]{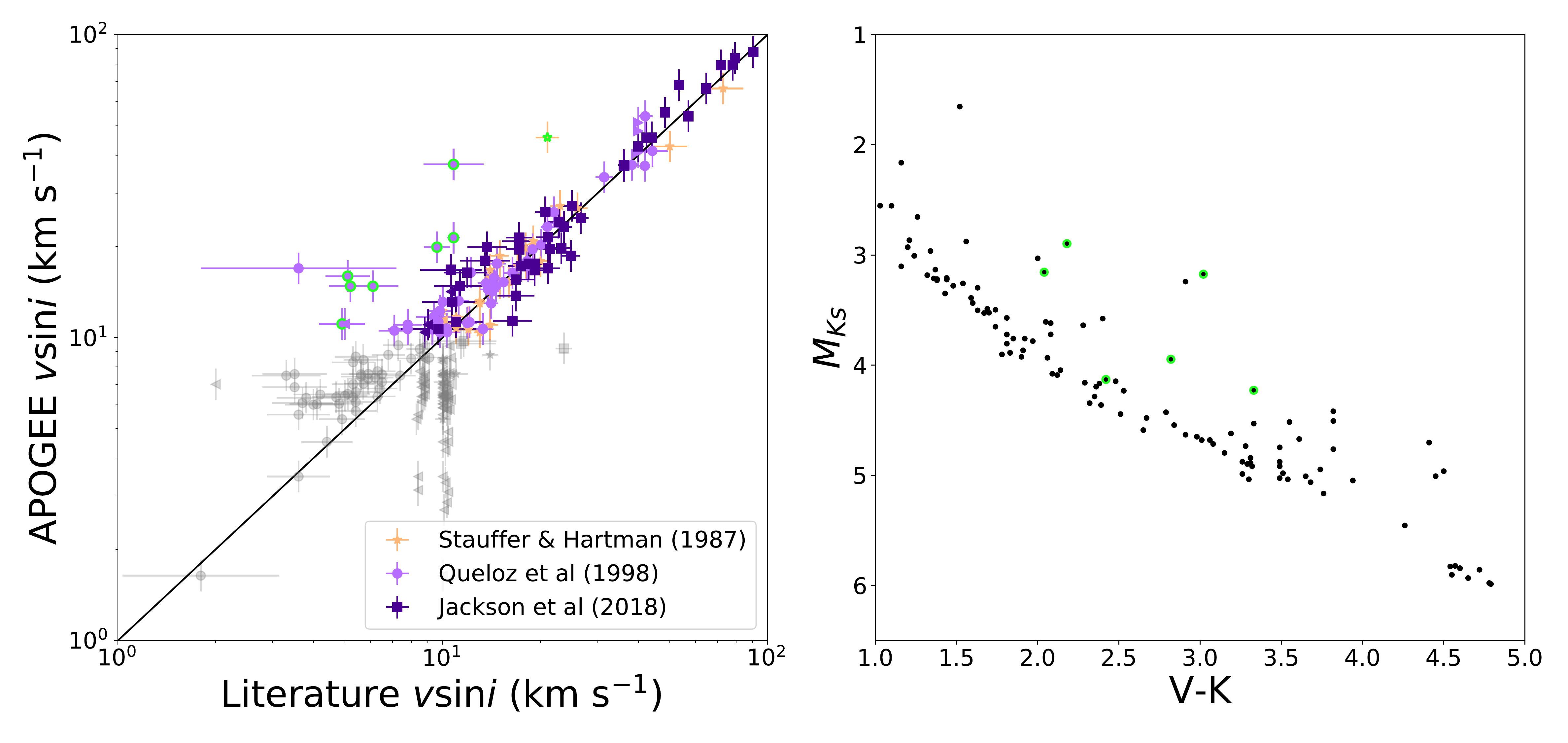}
  \caption{\emph{Left:} Comparison of APOGEE \vsini{} to the literature 
      \vsini{} for photometric singles from \citet{Stauffer87} (orange stars), 
      \citet{Queloz98} (violet circles), and \citet{Jackson18} (purple squares). 
      Grey points are below the APOGEE detection 
      limit of 10~\kms.  Left-pointing and right-pointing arrows are reported 
      as lower and upper limits in the original references. 8 outlying
      observations (for 7 unique targets) are denoted with a green outline. 
      \emph{Right:} The same targets in the \MK-(V-K) plane. The 7 targets with 
      outlying observations are denoted with a green outline, and almost all 
      lie on the photometric binary sequence.}
  \label{fig:vsinicomp}
\end{figure*}

We compare the APOGEE \vsini{} to literature \vsini{} in 
\cref{fig:vsinicomp}. We find that in all datasets, there is an apparent
increase in the variance between the APOGEE and literature \vsini{} at 
\(10~\kms\). We therefore adopt \(10~\kms\) as the minimum value for a robust 
detection. 

We determined the APOGEE uncertainty by modeling the literature and APOGEE
observations for the photometric single stars as independent observations with 
lognormally-distributed errors
for the \citet{Stauffer87}, \citet{Queloz98}, and \citet{Jackson18} samples. We
choose those three samples because there is a large overlap with APOGEE. 
We used an MCMC routine \citep{ForemanMackey13} with a likelihood function:
\begin{multline}
    L(\{v_{1,i}\}, \{v_{2,i}\} | \{\sigma_{1,i}\}, \sigma_2, \{v_{i}\}) =\\
    \prod_{i} \mathcal{N}(\log v_{1,i} ; \log v_i, \sigma_{1,i}) \times\\ \mathcal{N}(
    \log v_{2,i}; \log v_i, \sigma_2)
\end{multline}
where \(\{v_{1,i}\}, \{v_{2,i}\}\) and \(\sigma_{1,i}\) are the known
    literature \vsini{}, APOGEE \vsini{}, and literature \vsini{} error, 
    \(\{v_i\}\) are the unobserved true \vsini{}, and \(\sigma_2\) is the
    uncertainty we are trying to derive. We adopt flat priors for all
    \(\{v_i\}\) and \(\sigma_2\) between \(\log_{10}(1)\) and 
    \(\log_{10}(100)\). 

For the three datasets we used
for this analysis, the derived fractional uncertainties were 
\(12.2^{+2.0}_{-1.7}\), \(9.5^{+2.1}_{-1.7}\), and \(13.0^{+2.2}_{-1.9}\) 
As a result, we consider a reasonable uncertainty to be 12\%, about the order 
suggested by \citet{Tayar15} for APOGEE giants. 

In the whole sample we find a total of 7 outliers. We
identify outliers by calculating a \(\chi^2\) figure of merit between the
literature and APOGEE \vsini{} values for all targets with APOGEE
\(\vsini > 10~\kms\). We flag \vsini{} discrepant by \(3\sigma\) as 
inconsistent. 4 of the outliers were observed in both \citet{Queloz98} and
\citet{Jackson18}, and in each case the APOGEE value was discrepant with
\citet{Queloz98}, a high-resolution survey, and in agreement with 
\citet{Jackson18}, a medium-resolution survey. When we identify the
\vsini{} outliers in the HR diagram (see \cref{fig:vsinicomp}), we find that 
    six out of seven lie in the
photometric binary sequence. The one single outlier is the one which has an
APOGEE \vsini{} of \(11~\kms\), barely above the detection threshold. The
overall discrepancy rate is \(7/144=5\%\), and that for
photometric binaries is \(6/31=19\%\). The Pleiades data therefore confirms
that APOGEE detections for single stars on the lower main sequence are robust,
but that there may be additional systematics for photometric binary stars, which
will be more important for field samples.

\subsubsection{The Unevolved Lower Main Sequence}
\label{sec:cooldwarfs}

\begin{figure*}[htb]
  \centering
  \plottwo{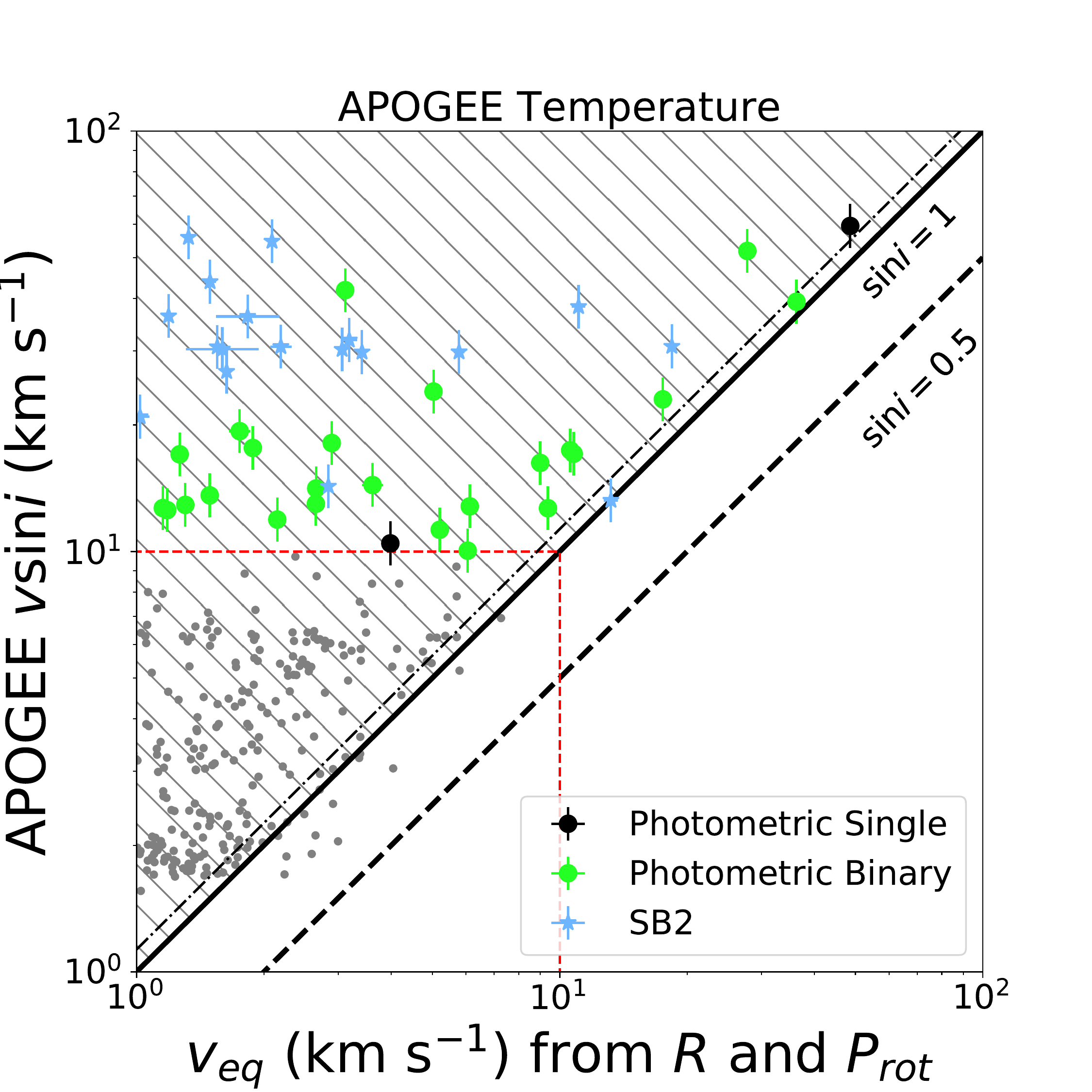}{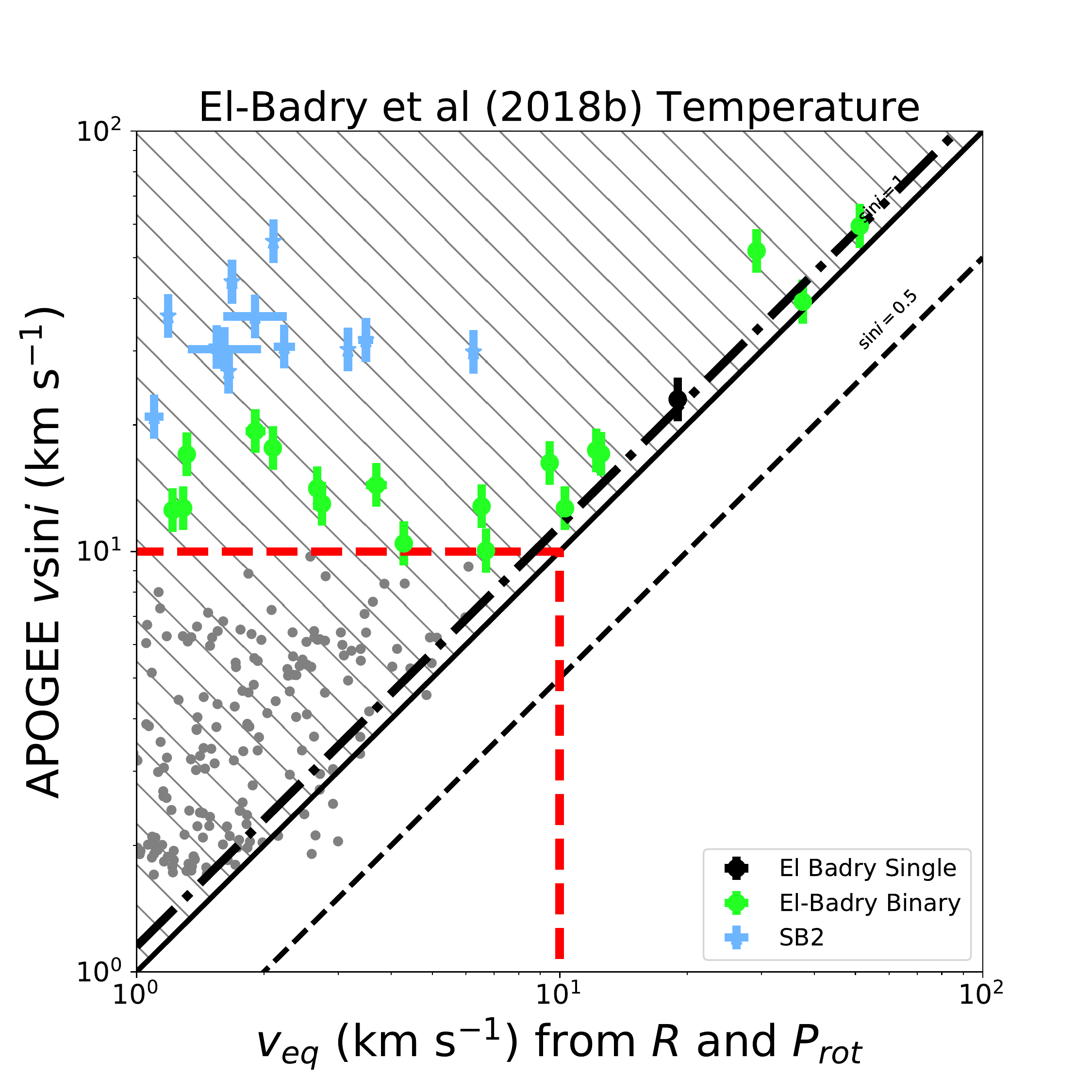}
  \caption{\emph{Left:} Same as \cref{fig:apokasc_vdists}, with photometric 
      binaries marked in green. The radius is determined by deprojecting 
      targets onto a MIST isochrone. A 13\% radius inflation is shown as the 
      dot-dashed line. \emph{Right:} Same as left except for the subset of 
      targets with corrected \Teff{} in \citep{ElBadry18b}. The radius is 
      recalculated using the corrected 
  \Teff.\label{fig:cool_stars_velocity_radius}}
\end{figure*}

Having compared directly to literature \vsini{}, we now turn back to the
\Kepler{} field where we validate against the \citet{McQuillan14} rotation
periods. Because binarity imposes a substantial uncertainty on
luminosity-derived radii, we restrict this validation to the unevolved lower
main sequence, where photometric binaries can be identified and their excess
luminosity corrected for through deprojection. In this sample, we have 512 
targets, 151 of which are photometric binaries. 

The comparison between the \vsini{} and equatorial velocity for the sample
overlapping with \citet{McQuillan14} is shown in the left panel of 
\cref{fig:cool_stars_velocity_radius}. We mark visual SB2s separately, and they 
behave as expected; the ASPCAP pipeline attempts to fit the two lines as a 
single peak and incidentally measures an unphysically high \vsini{}. 

However, even when these stars are not considered, there are striking
differences between the rapid rotator fractions measured by \vsini{}, and those
predicted from a combination of R and rotation period. The 
photometric binaries make up 25/27 of the rapid rotators (excluding SB2s), 
which is consistent with rapid rotation being associated with binaries 
\citep{Simonian19}. However, there is a large 
disagreement the spectroscopic and photometric rapid rotator fraction. The 
spectroscopic rapid rotator fraction
in the overlap sample is \(8.7^{+2.0}_{-1.7}\%\) while the predicted
photometric rapid rotator fraction is \(2.8^{+1.3}_{-0.9}\%\), a substantial
discrepancy.

Looking at the individual points, 24/27 of the spectroscopic rapid rotators 
have \vsini{} which is too large given the observed period and radius.
These outliers do not show clear signs of contamination as was found for the 
asteroseismic outlier.

This outlying sample separates into two classes which behave differently. There
is a population of stars with high \(v_{eq}\), but not as high as that
predicted by \vsini{}; and a slowly rotating population as gauged by
\(v_{eq}\), but with \vsini{} in the 10 to 15~\kms range. 

We are comparing
\vsini{} to a combination of rotation period and radius; the latter is deduced
from the measured \Teff{}-R relationship on a standard isochrone. There are
therefore three potential error sources: the period, the radius/temperature, or
the \vsini{}.

This discrepancy is unlikely to be due to the rotation period. While aliases
occur occasionally and may explain points off by a factor of two, the
\Kepler{} rotation periods have been extensively cross-validated
\citep{Aigrain15}. Additionally,
\citet{Simonian19} found that the rotation period distribution of the rapid
rotators matched the expected orbital period distribution of non-eclipsing
binaries inferred from the \Kepler{} eclipsing binary sample. Strong
systematics in the rotation periods would make this agreement unlikely. We are
therefore left with two possibilities: a problem in the radius, or a problem in
the \vsini.

\begin{figure}[htb]
    \centering
    \plotone{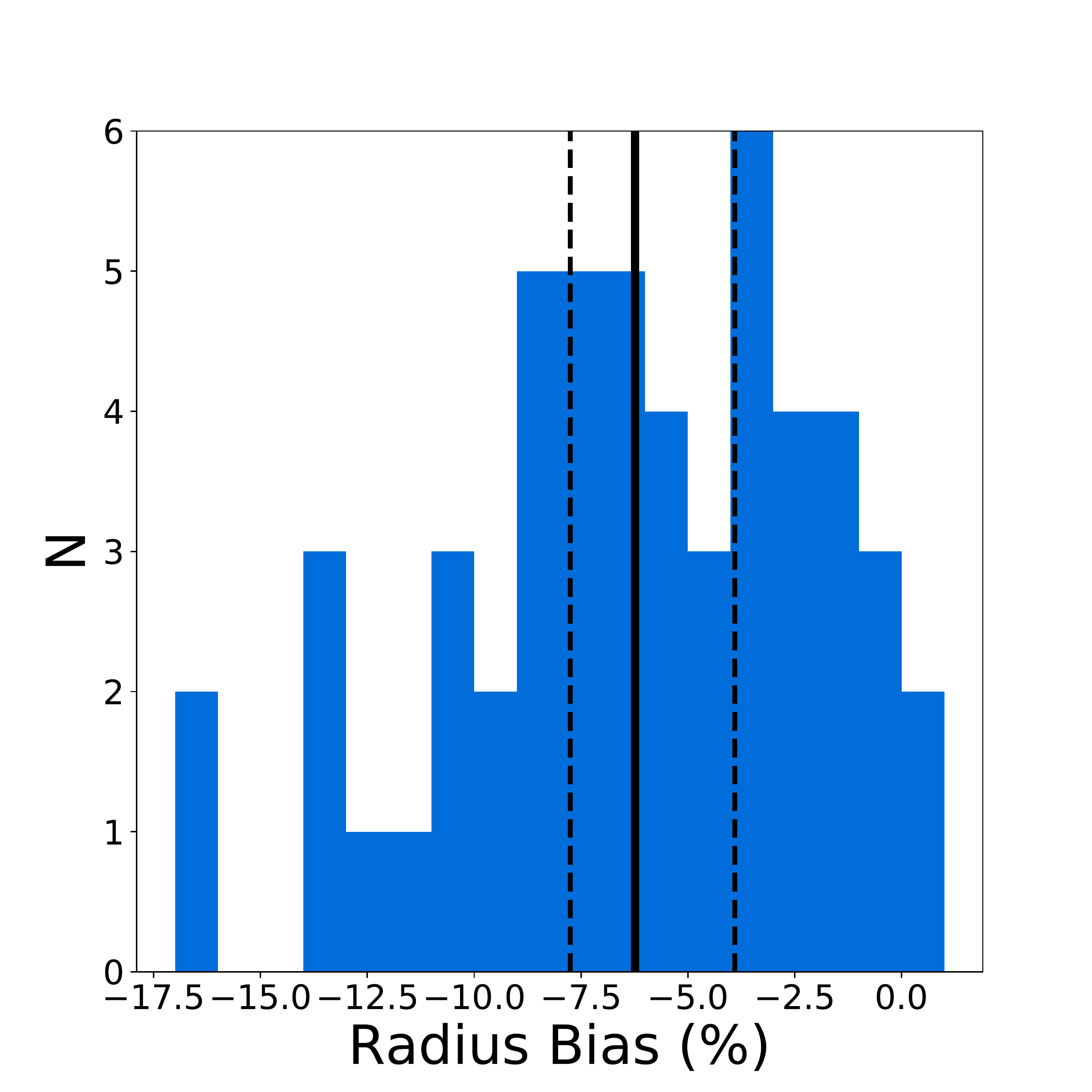}
    \caption{The distribution of radius underestimates for the sample of cool 
        dwarfs with decomposed primary \Teff{} measurements in 
        \citet{ElBadry18b}. The median underestimation is
    \(-5.9^{+2.0}_{-1.8}\), as indicated by the median and \(1\sigma\)
percentiles.}
    \label{fig:radiusbias}
\end{figure}

\citet{ElBadry18a} found that the APOGEE pipeline could underestimate the
\Teff{} of photometric binaries, which could produce biased radius estimates. 
The radius in this regime is determined from the
luminosity, bolometric correction, and temperature, all of which depend on the
adopted temperature due to the deprojection. To get a sense of the magnitude of
this effect, we calculate a radius bias using the subset of stars with
corrected temperatures in \citet{ElBadry18b}, which is shown in 
\cref{fig:radiusbias}.
The net effect of the temperature underestimation is that the radius is
underestimated by a median value of 5.9\%. To check if this improves the 
discrepancy seen in the cool dwarfs, we recompute \(v_{eq}\) using the revised 
radius in the right panel of \cref{fig:cool_stars_velocity_radius} for the 20 
non-SB2 targets.  While this does not explain the disagreement for the
low \(v_{eq}\) objects, this correction makes the high \(v_{eq}\) measurements
more plausible. 

We use standard isochrones to infer radii, which are known to underpredict the
true radii of active K and M stars \citep{Jackson14a}. Allowing for a radius 
inflation of 13\%, as shown in \cref{fig:cool_stars_velocity_radius}, can explain the
difference for stars which already have high \(v_{eq}\), roughly 1/5 of the
outliers. A different phenomenon is, however, required to explain the remaining
discrepancies.

Evidence that the majority of the high \vsini{} targets represent a background, 
rather than a true genuine rotation signal, is that most of these stars are 
either non-detection in
\citet{McQuillan14} or have long measured rotation periods. APOGEE rapid
rotators on the lower main sequence would be in a regime where
\citet{McQuillan14} would be very complete, and unlikely to either miss a
high-amplitude short period signal or see a low ampltiude long period signal
instead. Low \(\sin i\) might prevent 
detection in starspots, but would also cause low \vsini{}, not high.

We can compare the Pleiades and field star samples to quantify this bias. In 
the Pleiades, \(19^{+9.8}_{-7.4}\%\) of photometric binaries
have discrepant values. In the \Kepler{} field, the analogous fraction 
is \(12.5^{+3.3}_{-2.8}\%\), so the results from both populations are 
consistent with each other.

\subsection{Analysis}

The overall reliability of the APOGEE rotation data can be summarized as follows:

\begin{enumerate}
    \item \vsini{} and rotation periods are concordant for asteroseismic stars.
    \item \vsini{} and rotation periods are concordant for stars on the Pleiades single star sequence.
    \item \(19\%\) of photometric binaries in the Pleiades have \vsini{} which is too high for the rotation period.
    \item In all 4 discrepant Pleiades cases with both high- and medium-resolution literature \vsini{} values, the APOGEE \vsini{} agreed with the medium-resolution value, but was discrepant with the high-resolution value.
    \item \(13\%\) of photometric binaries on the unevolved lower main sequence in the \Kepler{} field have \vsini{} which are too high for the rotation period (there are too few rapid rotators on the single-star sequence to draw any meaningful conclusions).
    \item\label{item:rot} Of the \Kepler{} rapid rotators with discrepant rotation measures, around 1/5 have both \vsini{} and rotation period above the APOGEE \vsini{} detection threshold, while 4/5 have \vsini{} above the detection threshold, but the rotation period predicts that \vsini{} should be below the detection threshold.
    \item The temperature bias predicted by \citep{ElBadry18a} is not large enough to explain the discrepancy.
    \item Radius inflation can only explain the 1/5 of targets in \cref{item:rot}.
\end{enumerate}.

The combined corrections for radius
inflation for rapidly-rotating, active cool stars, and the effect of binarity 
on radius can resolve the 1/5 of cases, but not for the remaining discrepancies. 
We instead interpret the anomalously high APOGEE 
\vsini{} as the APOGEE spectrograph misclassifying unresolved binaries as rapid 
rotators, which can explain all the trends noted above. This would only occur for
binary systems for which the RV offset between the components is comparable to 
the velocity resolution of the spectrograph(\(\Delta RV = 
\textrm{c}/\mathcal{R} \sim 14~\kms\)). This mechanism would predict that there 
would be a sharp transition in \vsini{} between these spurious spectroscopic 
rapid rotators and resolved SB2s, which is evident from
\cref{fig:cool_stars_velocity_radius} at approximately twice the APOGEE
resolution. This mechanism would also explain the resolution-dependent
discrepancies seen in the Pleiades (see Section~\ref{sec:pleiades}), where the 
low-resolution instruments overestimate the \vsini{} compared to the 
high-resolution instruments.

This background is most visible in the cool dwarfs because the fraction of 
spurious rapid rotators \(19 / 512 = 3.7\%\) is large compared to the fraction 
of genuine rapid rotators in synchronized binaries, which is around 2\%
\citep{Simonian19}. As a result, a 4\% background becomes dominant in this
regime. 

\begin{figure}[htb]
    \centering
    \plotone{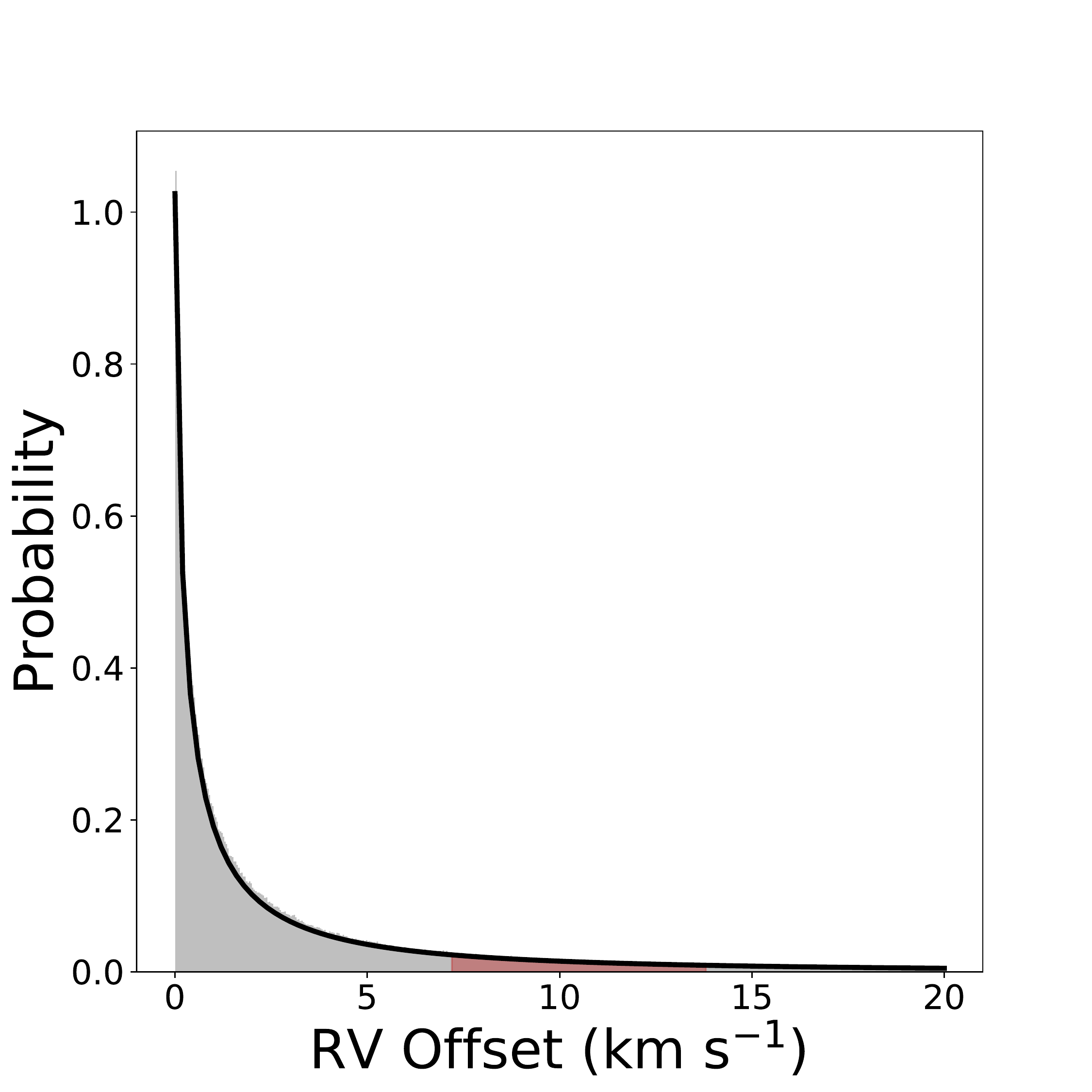}
    \caption{Distribution of RV offsets for a population of \(0.5 M_\sun\)
    equal-mass binaries in circular orbits following a \citet{Raghavan10}
    period distribution. A Monte-carlo realization is shown as the light gray
    histogram while an analytic expression is shown as the thick black line.
    The red region denotes the range between \(7~\kms < \Delta v < 14~\kms\),
    which is expected to correspond to the case of unresolved spectral blends.
    Around 10\% of the sample falls in between this range.
\label{fig:phot_cont}}
\end{figure}

To evaluate the plausibility of this mechanism, we make an order-of-magnitude 
estimate of how often this phenomenon should be observed in a
sample of photometric binaries. We build a toy model of a population of
equal-mass \(0.5 M_\sun\) binaries in circular orbits with a period 
distribution given by the lognormal distribution in \citet{Raghavan10}. We 
model the radial velocity difference of this population by drawing 1,000,000 
randomly distributed periods, inclinations, and orbital phase using Monte 
Carlo techniques. The histogram along with the analytic estimate is shown in
\cref{fig:phot_cont}. To estimate the fraction of spectral blends, we assume 
that all systems where the RV offset is between 0.5--1 times the resolution of 
the APOGEE spectrograph would correspond to spectral blends, which yields a
fraction of 10\% in our model photometric binary population. This value is 
comparable to the fraction that we see in real populations, indicating that 
this is a plausible
explanation for the phenomenon. More detailed modeling, outside of our scope,
would be required to quantify and remove this effect from the data set.

\section{Results}
\label{sec:results}

We've established that \vsini{} is a reliable tracer of rapid rotation, though
it contains a background of spuriously high \vsini{} targets associated with 
equal-luminosity binaries. To make scientifically interesting conclusions, we
seek populations where the signal of genuine rapid rotators is maximized with 
respect to this background. We argue
that the signal-to-background ratio is maximized for the rapidly rotating
subgiants.

As seen in \cref{fig:rapidmaps}, the genuine rapid rotator fraction in the
subgiant regime is high. The spectroscopic rapid rotator fraction in the regime
hotter than the fast-launch boundary predicted by \citet{vanSaders13}
is \(17.3^{+1.1}_{-1.0}\)\%, substantially higher than both the fractions
expected from spurious binaries and synchronized binaries. 

We also expect the background of
spectrally blended binaries to decrease for evolved stars because the 
luminosity contrast between binary components evolves with age. We have
three additional pieces of evidence that this effect becomes subdominant among 
the rapidly-rotating subgiants. First, based on \cref{fig:rapidmaps}, the rapid
rotator fraction clearly increases with luminosity, which is the opposite of
what would be expected from a background of near equal-mass binaries. 
Therefore, this background
must be small compared to the single-star rotation signal. Additionally, this
effect is not seen in the asteroseismic sample, which is not expected to have a
bias against binaries. 

Finally, we can see this effect diminish with age in the
background-limited regions of the HR diagram. If we remove the notch populated
by the rapidly-rotating subgiants, we find that the remaining sample 
demonstrates a transition (denoted by the aquamarine dotted line in
\cref{fig:rawdata}) between a high and low 
\vsini{} background. The low \vsini{} region (cool and luminous) has a
spectroscopic rapid rotator fraction of \(1.2^{+0.4}_{-0.3}\%\), which is 
consistent with the tidally-synchronized background. Based on the contours
shown in the left panel of \cref{fig:rapidmaps}, only a small fraction of the
tidally-synchronized population would undergo Roche-lobe overflow and merge. 
The high \vsini{} region (hot and less luminous) has a rapid rotator fraction 
of \(5.4^{+0.6}_{-0.5}\%\), as seen in the cool dwarfs, and is consistent with 
a 2\% synchronized binary population on top of a 4\% spectral blend population
measured in the cool dwarfs. We conclude that the effect of 
spectral blends is substantially decreased for evolved stars.

In the sample where \vsini{} is reliable, we note several interesting 
populations of rapid rotators. 

\subsection{Subgiant Rotation}
\label{sec:subgiants}

\begin{figure*}[htb]
    \centering
    \plottwo{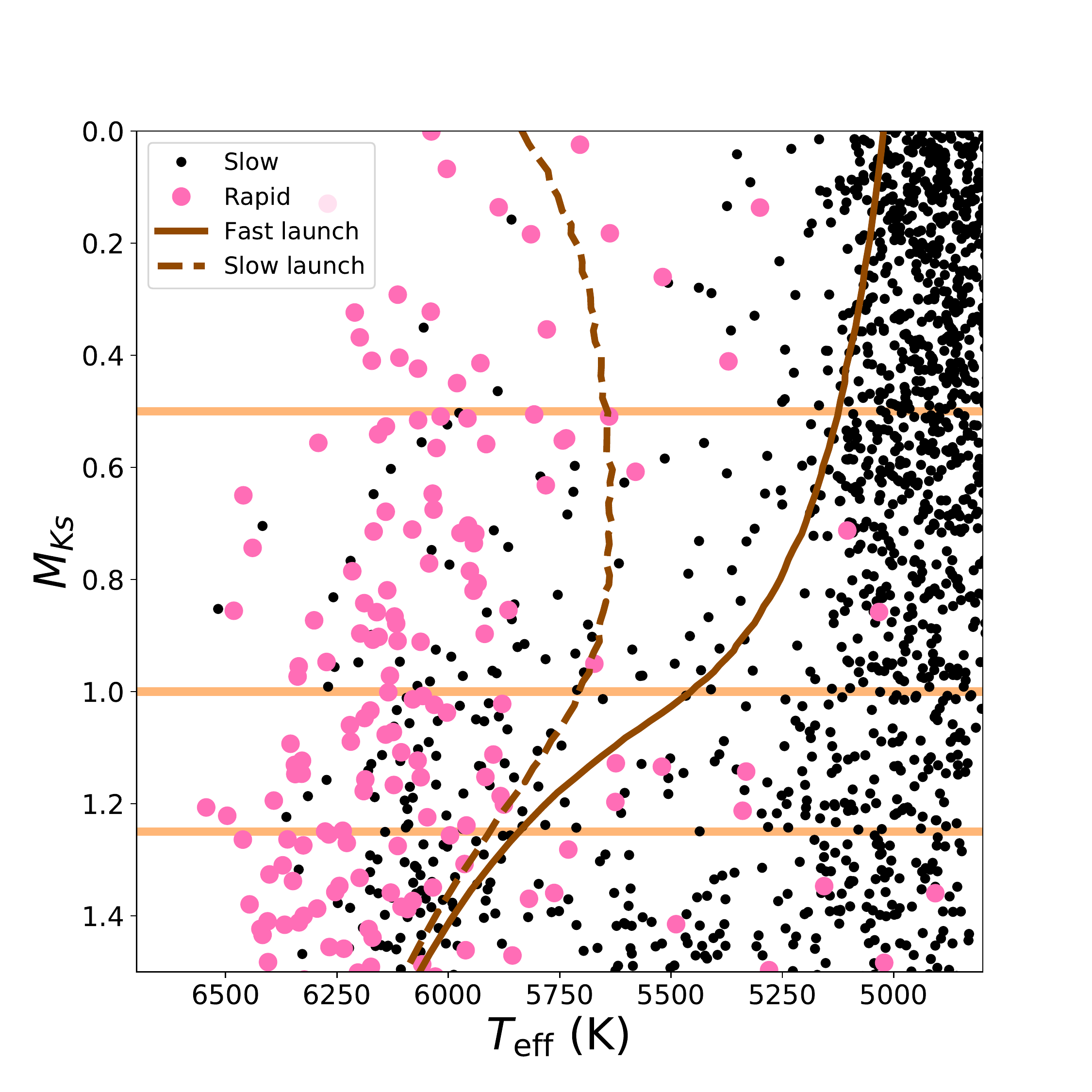}{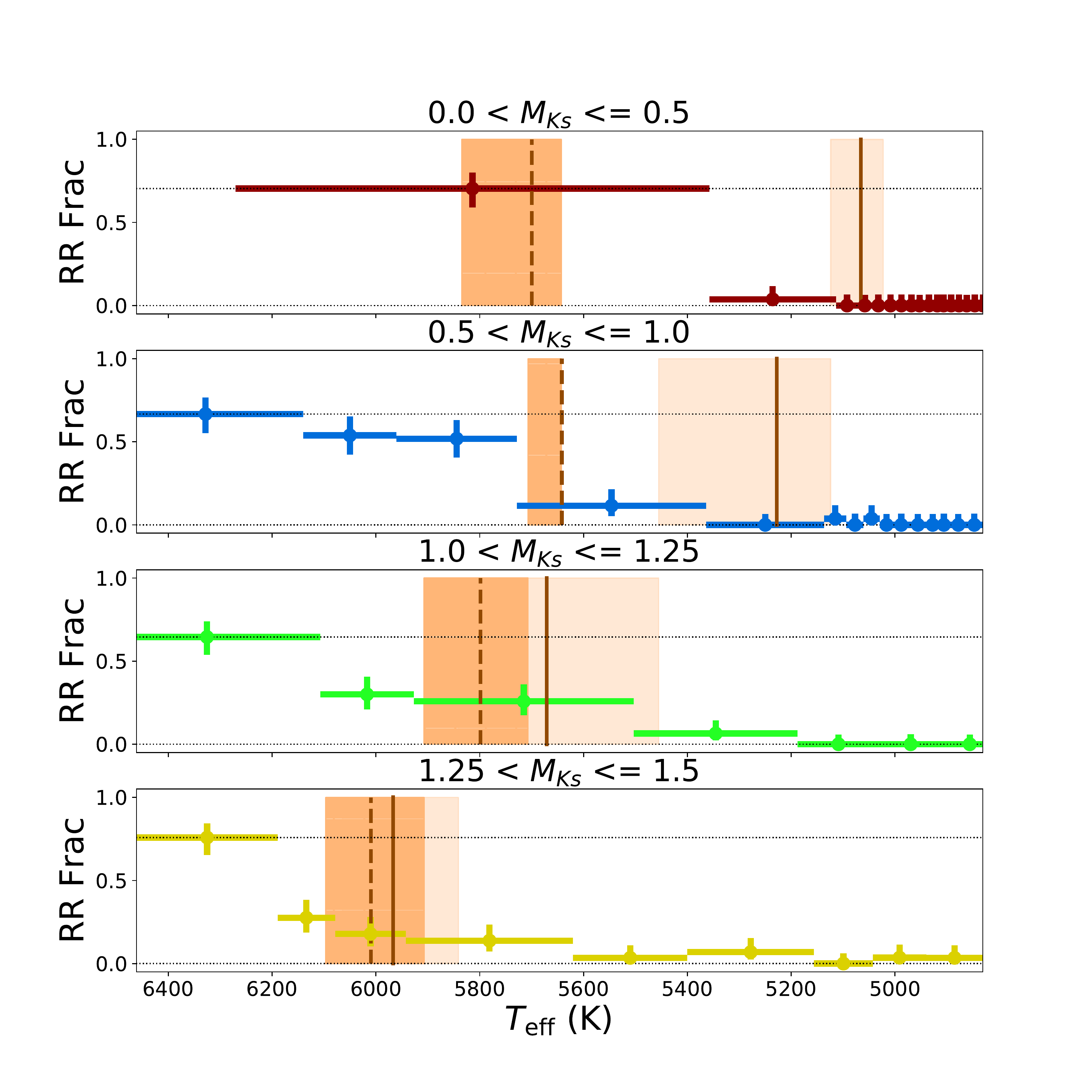}
    \caption{\emph{Left: } Rapid rotators emphasized in the zoom-in region
        denoted in the left panel of \cref{fig:rawdata}. Rapid rotators are
        denoted as large pink circles while the slow rotators are denoted as
        black dots. The orange horizontal lines illustrate the luminosity bins 
        for the right figure. The brown lines denote the boundaries predicted
        by \citet{vanSaders13}. \emph{Right:}  
        Rapid rotator fraction as a function of \Teff{} for four different
    bins in luminosity. The value of the slow and fast-launch model in the
center of each bin is shown as brown vertical lines (dashed and solid). The
range of model values within a bin is shown as the shaded region. The maximum
and minimum rapid rotator fractions are shown as a horizontal dotted lines.}
    \label{fig:rrbins}
\end{figure*}

We focus our model comparison for hot subgiants between \(2 \ge \MK > 0\) and \(6500 \textrm{ K } 
\ge \Teff{} > 4900 \textrm{ K }\) (shown as the orange box in
\cref{fig:rawdata}). We leave off the most luminous rapid rotators (delineated by the purple dashed lines in \cref{fig:rrbins}) because this region is too sparsely populated to make meaningful statistical comparisons of \vsini{} to models.

A zoom-in of the main sample is shown in the left panel of
\cref{fig:rrbins}. From just the raw data, we see that rapid rotators become
more sparse on the cool end. We plot predictions of the boundary separating detectable from
undetectable rotation for solar-metallicity stars \citep{vanSaders13}.
This ``iso-velocity'' boundary separates hot stars with \(v_{eq} > 10~\kms\) from 
cool stars below
our detection threshold. The fast and slow launch boundaries correspond to the 
fast and slow launch conditions described in \citet{vanSaders13}, and the
difference reflects the impact of the range of main sequence rotation rates on
the location of the \vsini{} cutoff.

We also note that a similar, albeit subdued trend is seen in 
the rotation period data in \cref{fig:rawdata}, despite 
hot subgiants not being expected to have substantial starspots. 

We perform a rough characterization of the spindown trends in
the right panel of \cref{fig:rrbins}, where the rapid rotation fraction with
temperature in bins of fixed luminosity is compared to that predicted by the
\citet{vanSaders13} models. For each luminosity bin, we 
plot the location of the boundary predicted by \citet{vanSaders13} for the slow 
and fast launch conditions, and the range of the models within each bin. As
expected, the detection fraction in the hottest stars is high, reaching a
maximum of 68\% for all bins; inclination effects and the presence of some very
slow rotators on the main sequence can explain why the detection fraction is 
not 100\% to the left of the slow launch boundary. In between the fast and slow 
launch boundaries, the rapid rotator fraction is expected to drop as
the upper envelope continues to decrease and the slower rotators become 
undetectable. Finally, all stars cooler than the
fast launch boundary should have undetectable rotation. The overall pattern is
close to that predicted by theory, confirming that the spindown models used for
main sequence stars are also viable when applied to hot subgiants evolving to
the red giant branch.

In the two 
least luminous bins, the rapid rotator fraction drops at cooler temperatures 
than that predicted by models. While this excess of rapid rotators could imply that
winds for stars near the Kraft break are overestimated, we note from
\cref{fig:rawdata} that the photometric binary sequence intersects the lowest
luminosity bins at \(\Teff \sim 6000\) K. We also note that a corresponding
increase in rapid rotators at that position in the HR diagram is not seen in 
the \citet{McQuillan14} rotation
data. The proximity of this excess to the photometric binary sequence, and its
disappearance in the rotation periods leads us to conclude that this 
discrepancy between the models and the data is likely not be the result of a 
deficiency in the model, but is rather caused by the high \vsini{} background 
observed near the photometric binary sequence.

\subsection{\(\alpha\)-rich Blue Stragglers}

\begin{figure}[htb]
    \centering
    \plotone{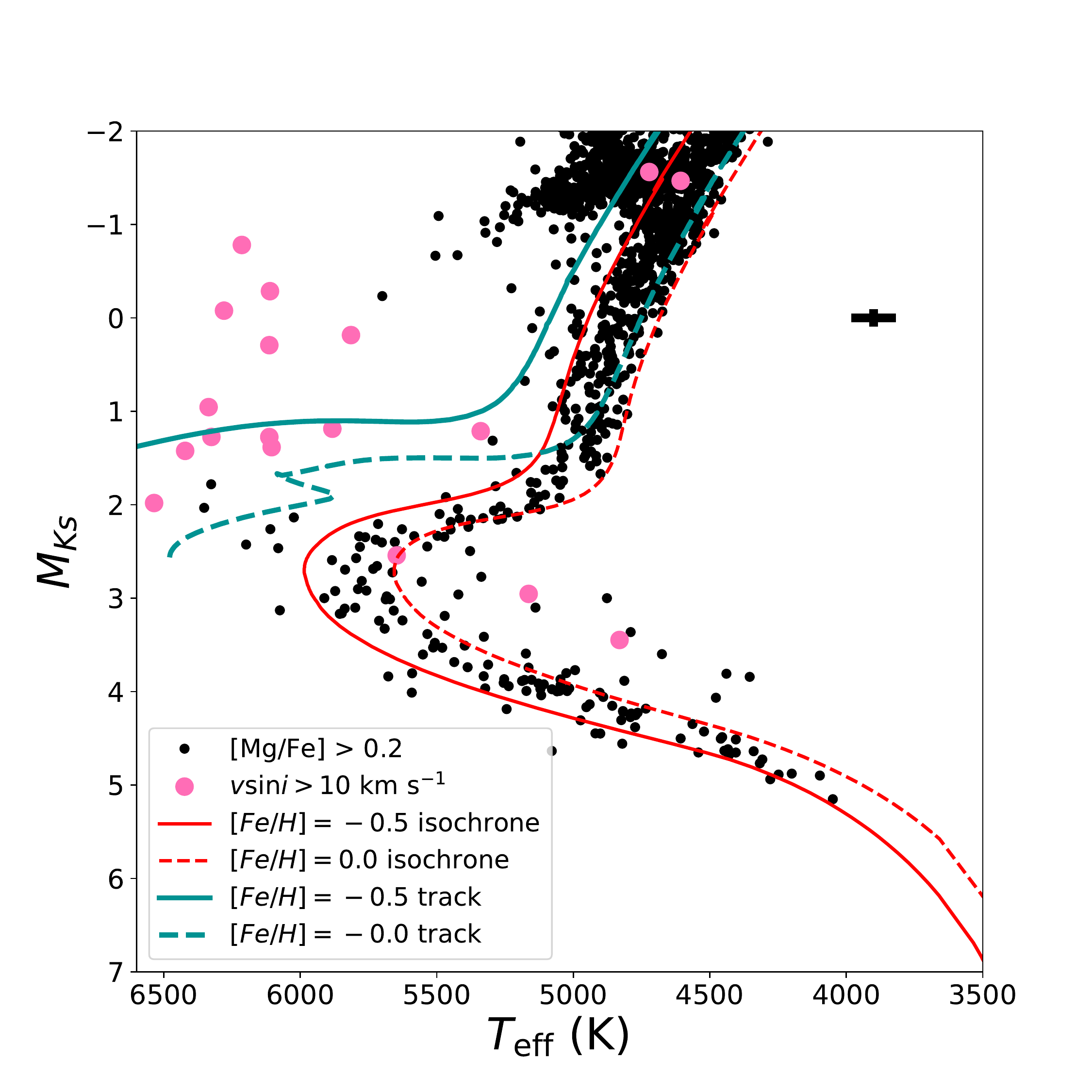}
    \caption{The \(\alpha\)-rich APOGEE sample with \(\mgfe > 0.2\). A
        representative error bar is shown in the upper right corner. The halo
        turnoff can be seen in this sample, between two 10~Gyr, \(\afe = 0.2\) 
        DSEP isochrones with  \(\feh = 0.0\) and \(\feh=-0.5\) (shown in red). 
        DSEP evolutionary tracks for stars representative of the Kraft break 
        (\(1.3 M_\sun\)), but with \(\afe = 0.2\) and \(\feh = -0.5, 0.0\), are
        shown in blue. Stars with \(\vsini > 10~\kms\) are shown as pink 
    circles.} 
    \label{fig:highalpha}
\end{figure}

The \Kepler{} field also contains a small population of \(\alpha\)-rich stars, 
which are generally an old population with a kinematic age of order 10 Gyr 
\citep{Schuster12}, well traced by \mgfe{} \citep{Hayes18}.  We isolate these 
stars by imposing a cut of \(\mgfe > 0.2\) and show them in 
\cref{fig:highalpha}.

In the high \(\alpha\) sequence, we find clear evidence of a turnoff,
which we trace between two alpha-enhanced (\(\afe = 0.2\)), 10 Gyr old DSEP isochrones: a
high metallicity one with \(\feh=0.0\), and a low metallicity one with \(\feh =
-0.5\). The stars hotter than the turnoff are blue stragglers.

Because we define rapid rotation at being at the 10 \kms{} level or higher, it is likely that all blue stragglers would form as rapid rotators by our criterion \citep{Sills05, Mucciarelli14}. However, only the stars hotter than the Kraft break would avoid severe spindown from magnetized winds. As a concrete example, \citet{Sandquist18} found clear evidence for a stellar merger product below the Kraft break on the M67 main sequence that we would count as a slow rotator. At solar metallicity, the Kraft break
occurs at 1.3 \(M_\sun\), and it is interesting to explore the metallicity
dependence of this phenomenon. So we
compare the location of the break in the high-\(\alpha\) sample to that
corresponding to \(\alpha\)-enhanced 1.3 \(M_\sun\) DSEP evolutionary tracks at
high and low metallicity in \cref{fig:highalpha}. We
find that the transition between rapid and slow rotators occurs about half-way
in between these two tracks, suggesting that the mass of stars at the Kraft
break does not change dramatically between solar abundance and this 
high-\(\alpha\) sample.

\subsection{Red Stragglers}
\label{sec:redstragglers}

\citet{Geller17} consolidated stars in numerous clusters which were redder than
the single-star isochrone and proposed a standardized nomenclature for them,
which we adopt here. ``Red Stragglers'' are defined as stars more luminous
than the turnoff which are substantially redder than the giant branch;
``Sub-subgiants'' are less luminous than the turnoff and redder than the main
sequence, placing them in between the main sequence and the subgiant branch
(see Fig. 1 in \citet{Geller17}).  
While the turnoff is not as well defined for a field population as it is in 
a cluster, we find likely analogs of red stragglers and sub-subgiants in the 
APOGEE field population. 

Both red stragglers and sub-subgiants occupy regions of the HR diagram which 
are not accessible to single-star evolution, and are believed to involve 
multi-star interactions. The three most promising explanations are that
the red stragglers are: (1) mass-transfer systems out of thermodynamic 
equilibrium, (2) stripped subgiants from close encounters, and (3) 
underluminous, magnetically active subgiants in a synchronized binary 
\citep{Leiner17}. All three mechanisms would leave a signature of rapid 
rotation as a result of the stellar interaction, which we see in the APOGEE
data.

In the red stragglers, we find definite signatures of rapid rotation. Of the 7
red stragglers in our sample, 6 have \(\vsini > 10~\kms\), which is consistent
with a fully rapidly-rotating population. Of this
sample, 3 have been observed by APOGEE over multiple epochs, and all show RV 
variability greater than \(30~\kms\), demonstrating that they are likely close 
binaries. One red straggler has a rotation period of 10.8 days, which is 
consistent with its radius and \vsini{}, and also consistent with estimates of the
synchronization threshold on the main sequence \citep{Mazeh08}. These systems
are extremely prominent in the HR diagram, and are likely an interesting
population to study in the field with APOGEE.

Of the 28 stars in the sub-subgiant region, only a few appear to be true 
sub-subgiants. 6 of them are visual SB2s, which we exclude because their stellar
properties are likely unreliable. 7 of the remaining 
sample have \(\vsini > 10~\kms\) including 2 lower limits, implying a
rapid rotation fraction of 32\%. Two of the spectroscopic rapid rotators have 
rotation periods in \citet{McQuillan14}. One has a 23 day period which is too 
long for its \vsini{}, demonstrating the presence of spectral blends in this 
sample. The other has
a 0.3 day period and an upper limit of \(\vsini > 87~\kms\), implying it's a
genuine rapid rotator. However, its does not have detectable RV variability,
which means it may not have a close binary companion, and would be an
interesting target for further study.  We conclude that while this regime does 
contain a genuine sample of rapid rotators, biased stellar parameters
caused by binarity can also spuriously populate this
region. We suggest that special care be taken to study these targets in detail
with specifically chosen methods that are robust to binarity.

\section{Discussion and Conclusion}
\label{sec:discussion}

We validated the catalog \vsini{} for a sample of 5,337 dwarfs and subgiants
observed by APOGEE by comparing against
Kepler rotation periods and literature \vsini{}. In this sample, we found: 

\begin{itemize}
    \item Moderate-resolution spectroscopy produces a spuriously high
        \vsini{} background caused by spectral blending of binaries at the
        level of 4\% of the population.
    \item This background is less common for evolved populations. It can be
        negligible for intrinsically rapidly rotating populations such as the
        subgiants, for which we found agreement with the models of
        \citet{vanSaders13}.
    \item The location of the Kraft break for a high-\(\alpha\) population is
        consistent with that expected based on the solar-metallicity
        population.
    \item APOGEE can detect rapid rotation for exotic classes of synchronized
        binaries such as sub-subgiants and red stragglers.
\end{itemize}

\citet{Huber14} enlarged the sample of stars with KIC parameters by using 2MASS
data for objects without \textit{griz} photometry. The agreement between their
properties and those in APOGEE is reasonable (see Appendix~\ref{sec:huber} for
details), but we caution that the agreement between the photometric and
spectroscopic data is degraded relative to the performance in the sample with
\textit{griz} data.

In the age of large datasets, binarity needs to be adequately corrected for in
order to measure the effects of rare populations. With the advent of
\Gaia{} parallaxes, luminosity information is widely available as a useful 
tool to separate single stars from photometric binaries. For analyses focusing 
on single stars, removing the photometric binaries can be a 
useful step in ensuring a well-behaved sample despite the fact that it will 
remove genuine hot evolved stars. Modeling is another technique to 
account for the effects of binaries. However, it is complicated by the fact that 
the binary population is highly sensitive to how the sample was selected. For 
example, binaries will be overrepresented in magnitude-limited samples 
\citep{Simonian19} so their impact on the population will necessarily change 
depending on the sample.

\begin{figure*}[htb]
    \centering
    \plotone{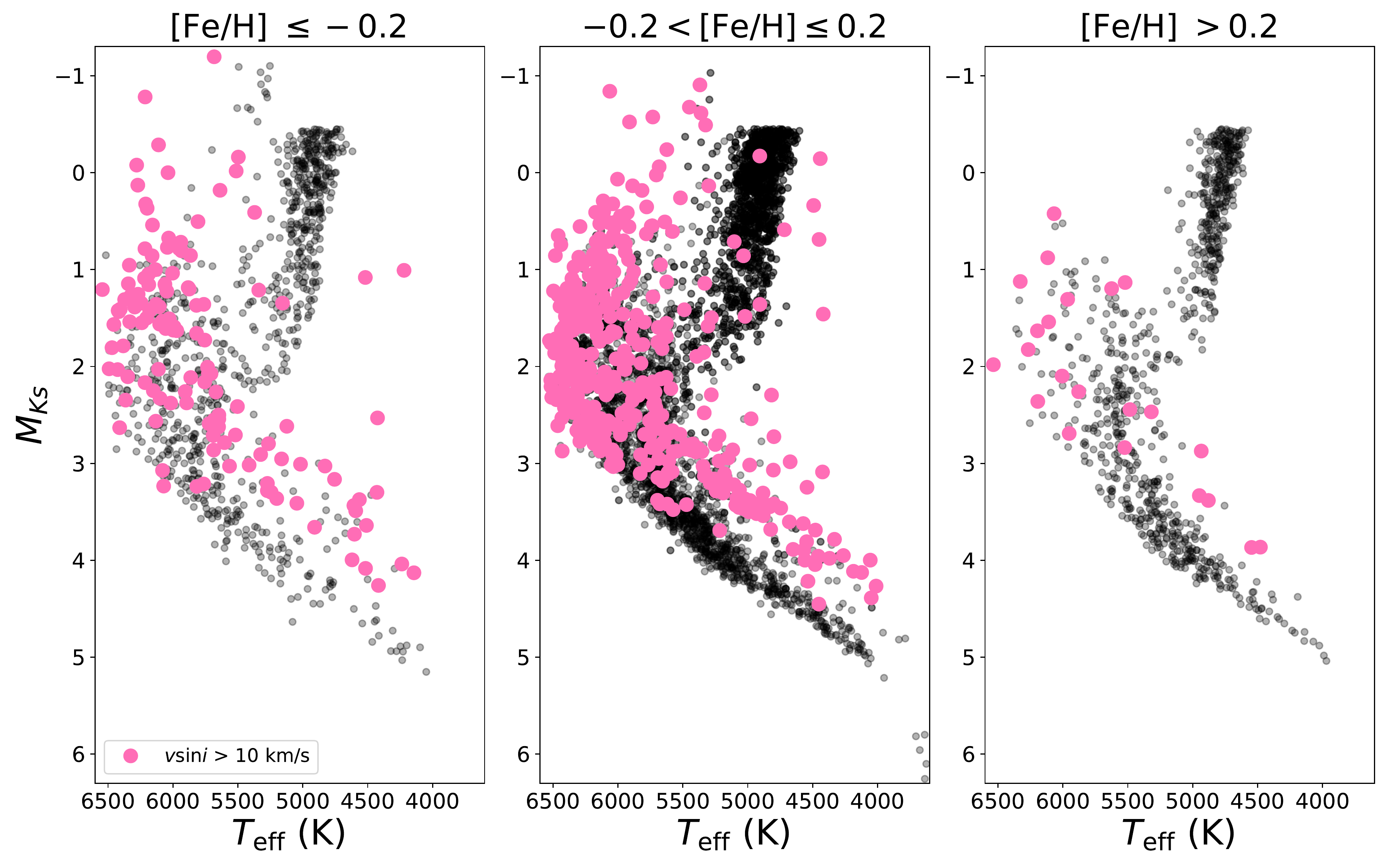}
    \caption{Rapid Rotators highlighted for three bins of metallicity with
        boundaries of \(\feh=-0.2, 0.0, 0.2\). Rapid rotators are emphasized as
        pink circles.}
    \label{fig:metallicity}
\end{figure*}

We touched upon the largely unstudied role of metallicity on rotation in the
high-\(\alpha\) sample, and explore metallicity more fully by binning the sample by metallicity in 
\cref{fig:metallicity} with the corresponding metallicity-adjusted 
\citet{vanSaders13} models. There are qualitative differences in the behavior 
of the model and the data, but our sample does not have enough metal-rich and 
metal-poor stars to meaningfully constrain the behavior of the models. One way 
to increase the number of high and low-metallicity targets is to expand the sample 
beyond the \Kepler{} field, which can be used to further constrain the behavior 
of rotation with metallicity.

Investigations into metallicity trends with rotation need to be wary of trends
in binarity with metallicity. Recent results have shown that the field binary
fraction may be anticorrelated with metallicity \citep{ElBadry19,Moe19}. These
results are in agreement with what we observe in the Kepler field, shown in
\cref{fig:metallicity}. The metal-poor regime contains a large fraction
of photometric binaries (hence high-vsini stars), while the metal-rich regime
is nearly devoid of photometric binaries. Investigations into how a population
which contains binaries changes with metallicity will need to control for these
effects.

\acknowledgments

G.S, M.P. and D.T acknowledge support from NASA ADP Grant NNX15AF13G and from
the National Science Foundation via grant AST-1411685 to The Ohio State
University. G.S. is also grateful to Jamie Tayar for illuminating conversations
regarding the APOKASC catalog as well as stellar rotation in general. We also
thank the anonymous referee for their highly clarifying suggestions.

\facility{Kepler, Gaia, CTIO:2MASS, Sloan}

\software{MIST \citep{Choi16}, Astropy \citep{astropy}, IPython \citep{PER-GRA:2007}, Scipy
\citep{jones_scipy_2001}, NumPy \citep{van2011numpy}, Matplotlib \citep{Hunter:2007} }

\bibliographystyle{aasjournal}
\bibliography{references}

\appendix
\section{Huber et al (2014) temperature}
\label{sec:huber}

One significant benefit of the APOGEE targeting criteria being based on
\citet{Huber14} stellar parameters is that APOGEE independently measured
temperatures for stars which previously only had \Teff{} and evolutionary 
state classifications from 2MASS colors. We perform a quick validation of 
those parameters to determine the reliability of stellar classifications based 
only on 2MASS photometry.

\begin{figure*}[htb]
    \centering
    \epsscale{1.2}
    \plotone{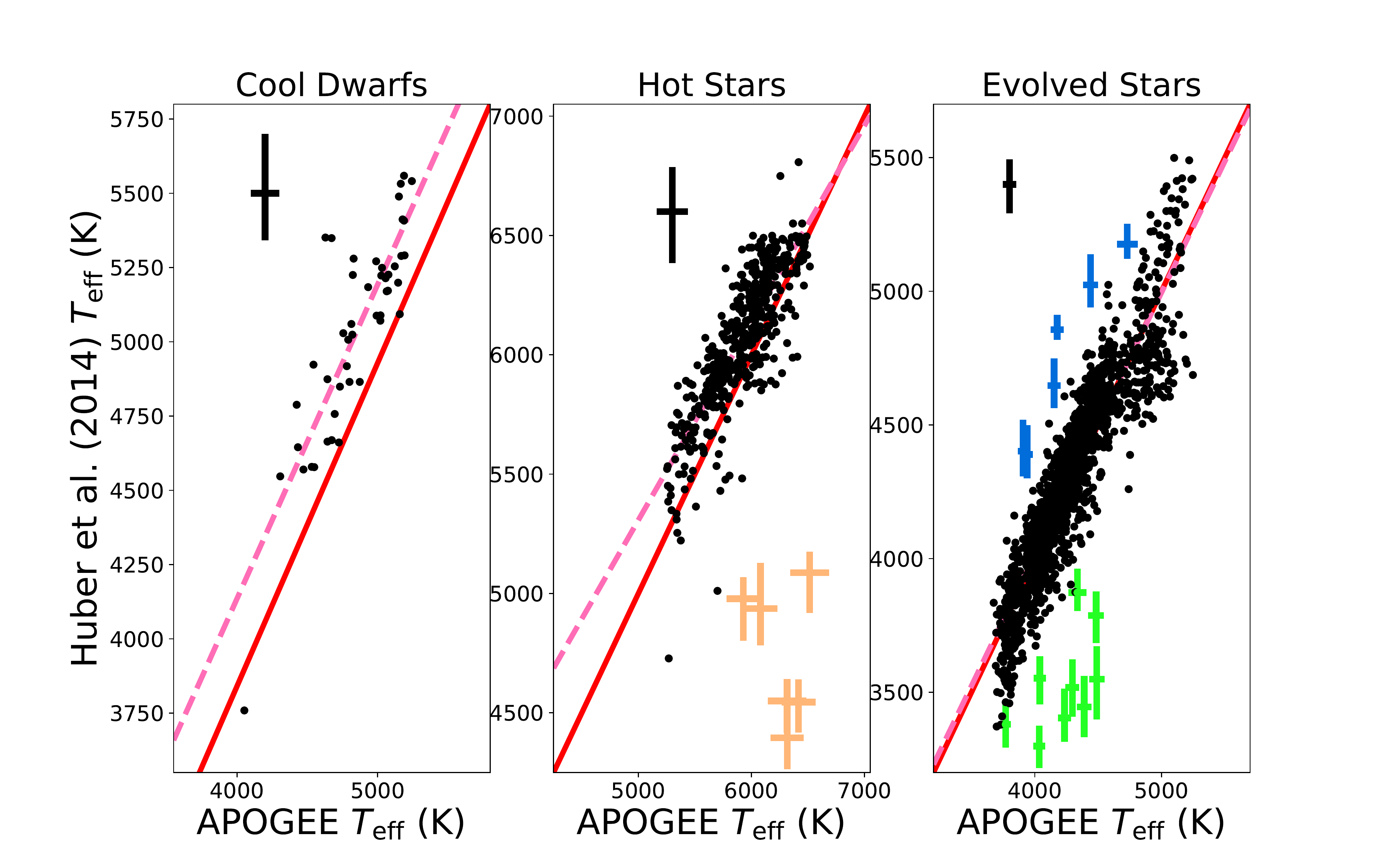}
    \caption{\emph{Left:} Comparison between \citet{Huber14} and APOGEE
    effective temperatures for cool dwarfs and photometric binaries with 
    temperatures determined from
    2MASS photometry and independently measured by APOGEE\@. A representative
    error bar for the bulk of the sample is shown in the top left corner. The 
    solid red line denotes the one-to-one relation. The dashed pink line 
    denotes the best linear fit with equation \(\Teff \textrm{(Huber)} = 1.06
\Teff \textrm{(APOGEE)} - 98 \textrm{ K}\) with a scatter of 177 K. 
\emph{Middle:} Same as left, except for the hot star sample. The 
best-fit relation is given as \(\Teff \textrm{(Huber)} = 0.83 \Teff 
\textrm{(APOGEE)} - 1173 \textrm{ K}\) with an RMS scatter of 159 K. Points 
with APOGEE \Teff{} more than \(4\sigma\) higher than the one-to-one relation 
are marked as orange points, and were excluded from the fit and scatter. 
\emph{Right:} Same as left, except for the evolved stars. The best-fit 
relation in this regime is given as \(\Teff \textrm{(Huber)} = 0.98 \Teff 
    \textrm{(APOGEE)} - 91 \textrm{ K}\) with a scatter of 142 K. Points with 
    APOGEE \Teff{} more than \(4\sigma\) greater than the one-to-one line are 
    marked in green, while points with APOGEE \Teff{} more than \(4\sigma\) 
    lower than the one-to-one line are marked in blue. Both sets of outliers 
were excluded from the fit and scatter.}
    \label{fig:huber_comparison}
\end{figure*}

\begin{figure*}[htb]
    \centering
    \plotone{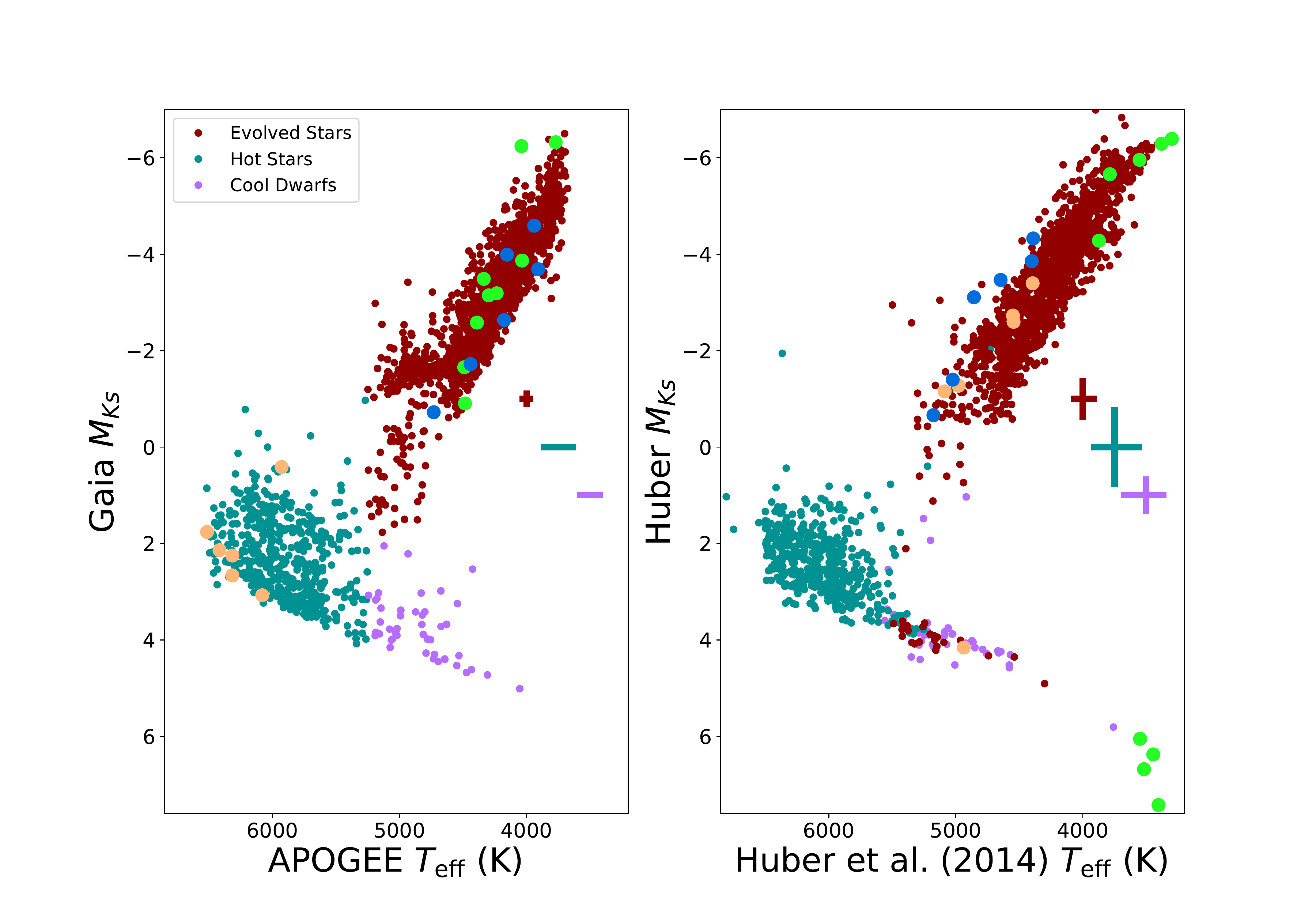}
    \caption{\emph{Left:} Standard evolutionary state classifications using the
    APOGEE and \Gaia{} parameters in this work compared to those inferred by
    \citet{Huber14}. Outliers from \cref{fig:huber_comparison} are plotted as 
    orange, green, and blue points.  Median error bars for each class are 
    shown in the right-hand side. \emph{Right:} Same as left, except using 
    stellar parameters derived from 2MASS colors by \citet{Huber14}.}
    \label{fig:hub_class}
\end{figure*}

In \cref{fig:huber_comparison}, we plot APOGEE \Teff{} against the 2MASS-based 
temperature calculated by \citet{Huber14} for three regimes: 
cool dwarfs, hot stars and evolved stars. These regimes were defined coursely:
the cool dwarfs and evolved stars have APOGEE \(\Teff < 5250\) K and 
\(\MK > 2\) and \(\MK \le 2\) respectively, the 
hot stars have APOGEE \(\Teff \ge 5250\) K. The classifications are illustrated
in the left panel of \cref{fig:hub_class}. We flag stars with temperatures 
differing by \(4\sigma\) as outliers, and exclude them from the best-fit 
relations calculated in the comparison.

We find that the agreement between \citet{Huber14} and APOGEE temperatures 
depends highly HR diagram position. For cool dwarfs, the slope of the 
temperature
relation is nearly unity, but the \citet{Huber14} temperatures are hotter than
the APOGEE temperatures by 98 K. For the hot stars, the 
relationship between \citet{Huber14} and APOGEE temperatures is more complex, as the slope is
substantially different than one, likely reflecting that the 2MASS temperatures
perform more poorly for the hotter stars, which is expected. We also note the 
presence of a small clump of 
six outlying points, plotted in orange. For the giant stars, 
both the slope and intercept are statistically indistinguishable from the 
one-to-one line, but there is a substantial residual corresponding to the red 
clump, which is resolved by APOGEE but not by \citet{Huber14}. In this regime
we flag two sets of outliers, a set 
of six where the \citet{Huber14} temperature is greater than APOGEE (shown in
blue), and a set of nine outliers where the APOGEE temperature is higher
(shown in green). Of the 2,043 stars in the overlap sample between 
\citet{Huber14} and APOGEE, only 21 are inconsistent outside the errors, or 
about a 1\% discrepancy rate.

We also compare the evolutionary state classifications for \Gaia{} and APOGEE
to those inferred by \citet{Huber14} using 2MASS colors. We estimate the
\citet{Huber14}-inferred \Ks-band absolute magnitude by substituting the
\citet{Huber14} radius into the Stefan-Boltzmann law and applying a bolometric
correction consistent with the \citet{Huber14} \Teff. In \cref{fig:hub_class},
we find that gross misclassifications are extremely rare.  First, we note 
that the orange outliers correspond directly to misclassified hot stars. We 
find that 5/1486 or 0.3\% of 
\citet{Huber14}-classified giants are grossly misclassified hot stars. There 
are no gross misclassifications of either cool stars or giants in the hot star 
domain. For stars whose \citet{Huber14} parameters would classify them as cool
dwarfs, 31/48, or 64\%, of these stars are misclassified giants, which is to be
expected as dwarfs overlap with giants in 2MASS colors.
However, this disagreement is exaggerated by the dearth of cool stars in this
sample compared to the full \Kepler{} sample, and does not extend to the 
temperature estimates. Only 4 of the 31 misclassified red giants
have 2MASS temperature estimates that differ substantially from the APOGEE
temperature. We finally note that the remainder of the
green outliers and all of the blue outliers have consistent evolutionary state
classifications as giants, but simply have discrepant temperatures, indicating
an outlier rate of 9/1486, or 0.6\%.

\section{Pleiades Cross-matched Sample}

The cross-matched sample of Pleiades targets in given in \cref{tab:pleiadessample} below.

\movetabledown=3in
\begin{rotatetable*}
\begin{deluxetable*}{ l l c c c c c c c c c c c c}
\tablecaption{Pleiades Overlap Sample\label{tab:pleiadessample}}
\tablehead{\colhead{} & \colhead{} & \multicolumn{3}{c}{\citet{Stauffer87}} & \multicolumn{2}{c}{\citet{Queloz98} ELODIE} & \multicolumn{3}{c}{\citet{Queloz98} CORAVEL} & \multicolumn{3}{c}{\citet{Jackson18}}\\ 
\cmidrule(lr){3-5}\cmidrule(lr){6-7}\cmidrule(lr){8-10}\cmidrule(lr){11-13} 
\colhead{APOGEE ID} & \colhead{Other ID} & \colhead{Limit} & \colhead{$v \sin i$} & \colhead{Error} & \colhead{$v \sin i$} & \colhead{Error} & \colhead{Limit} & \colhead{$v \sin i$} & \colhead{Error} & \colhead{Limit} & \colhead{$v \sin i$} & \colhead{Error}\\ \colhead{ } & \colhead{ } & \colhead{ } & \colhead{$\mathrm{km\,s^{-1}}$} & \colhead{$\mathrm{km\,s^{-1}}$} & \colhead{$\mathrm{km\,s^{-1}}$} & \colhead{$\mathrm{km\,s^{-1}}$} & \colhead{ } & \colhead{$\mathrm{km\,s^{-1}}$} & \colhead{$\mathrm{km\,s^{-1}}$} & \colhead{ } & \colhead{$\mathrm{km\,s^{-1}}$} & \colhead{$\mathrm{km\,s^{-1}}$}}
\startdata
2M03425511+2429350 & HII 25 &  &  &  &  &  &  & 44.2 & 4.9 &  &  &  \\
2M03430293+2440110 & HII 34 &  &  &  & 5.9 & 0.8 &  & 7.3 & 1.1 & < & 8.7 &  \\
2M03433195+2340266 & HII 120 &  & 11.0 & 0.6 &  &  &  & 9.4 & 1.2 &  &  &  \\
2M03433659+2327142 & HII 146 & < & 10.0 &  &  &  &  &  &  &  & 16.4 & 2.3 \\
2M03433662+2413562 & HII 134 &  & 65.0 &  &  &  &  &  &  &  &  &
\enddata
\tablecomments{Columns: (1) The 2MASS designation for the target. (2) Other designations which are used in the original \vsini{} papers. (3), (4), and (5) represent the limit designation, \vsini{} and uncertainty reported in \citet{Stauffer87} (6) and (7) represent the \vsini{} and uncertainty reported in \citet{Queloz98} with the ELODIE instrument. (8), (9), and (10) represent the limit designation, \vsini{}, and uncertainty reported in \citet{Queloz98} with the CORAVEL instrument. Finally, (11), (12), and (13) represent the limit designation, \vsini{} and uncertainty reported in \citet{Jackson18}. \Cref{tab:pleiadessample} is published in its entirely in the machine-readable format. A portion is shown here for guidance regarding its form and content.}
\end{deluxetable*}

\end{rotatetable*}

\end{document}